\documentclass[twocolumn,showpacs,aps,epsfig,nofootinbib]{revtex4}

\usepackage{graphicx}
\usepackage{epstopdf}
\usepackage{latexsym}
\usepackage{amssymb}
\usepackage{amssymb}

\usepackage[center]{subfigure}

\begin{document}

 \newcommand{\bq}{\begin{equation}}
 \newcommand{\eq}{\end{equation}}
 \newcommand{\bqn}{\begin{eqnarray}}
 \newcommand{\eqn}{\end{eqnarray}}
 \newcommand{\nb}{\nonumber}
 \newcommand{\lb}{\label}
\newcommand{\PRL}{Phys. Rev. Lett.}
\newcommand{\PL}{Phys. Lett.}
\newcommand{\PR}{Phys. Rev.}
\newcommand{\CQG}{Class. Quantum Grav.}

\title{General covariant Horava-Lifshitz gravity without projectability condition and its applications to cosmology}

\author{Tao Zhu${}^{a}$}
\email{zhut05@gmail.com}

\author{Fu-Wen Shu ${}^{b}$}
\email{shufw@cqupt.edu.cn}

\author{Qiang Wu ${}^{a}$}
\email{wuq@zjut.edu.cn}

\author{Anzhong Wang ${}^{a, c}$}
\email{anzhong_wang@baylor.edu}

\affiliation{${}^{a}$ Institute  for Advanced Physics $\&$ Mathematics,   Zhejiang University of
Technology, Hangzhou 310032,  China \\
${}^{b}$ College of Mathematics and Physics, Chongqing University of Posts and Telecommunications, Chongqing 400065, China\\
${}^{c}$ GCAP-CASPER, Physics Department, Baylor
University, Waco, Texas 76798-7316, USA }

\date{\today}

\begin{abstract}
We consider an extended theory of Horava-Lifshitz gravity with the detailed balance condition softly breaking, but without the projectability
condition. With the former, the number of independent coupling constants is significantly reduced. With the latter and by extending the
original foliation-preserving diffeomorphism symmetry $ {\mbox{Diff}}(M,  {\cal{F}})$ to include a local $U(1)$ symmetry,  the spin-0
gravitons  are eliminated. Thus, all the problems related to them disappear, including the instability,  strong coupling, and different speeds
in the gravitational sector. When the theory couples to a scalar field, we find that the scalar field is not only stable in both the ultraviolet (UV)
and infrared (IR), but also free of the strong coupling problem, because of the presence of  high-order spatial derivative terms of the scalar
field. Furthermore, applying the theory to cosmology,  we find  that due to the additional $U(1)$ symmetry, the Friedmann-Robertson-Walker
(FRW) universe is necessarily flat. We also investigate the scalar, vector, and tensor perturbations  of the flat FRW universe, and derive the
general  linearized field equations for each kind of  the perturbations.
\end{abstract}


\maketitle


\section{Introduction}
\renewcommand{\theequation}{1.\arabic{equation}} \setcounter{equation}{0}

Recently, Horava formulated a theory of quantum gravity,   whose scaling at short
distances exhibits a strong anisotropy between space and time  \cite{Horava},
\bq
\lb{1.1}
{\bf x} \rightarrow b^{-1} {\bf x}, \;\;\;  t \rightarrow b^{-z} t.
\eq
 In order for the theory to be power-counting
renormalizable,   in $(3+1)$- dimensions the critical exponent $z$   needs to be $z \ge 3$ \cite{Horava,Visser}.
The gauge symmetry of the theory now is broken from the general covariance, $\tilde{x}^{\mu} = \tilde{x}^{\mu}(t, x)\; (\mu = 0, 1, 2, 3),$
down to  the  foliation-preserving
diffeomorphisms, $\mbox{Diff}(M, \; {\cal{F}})$,
\bq
\lb{1.4}
\tilde{t} = t - f(t),\; \;\; \tilde{x}^{i}  =  {x}^{i}  - \zeta^{i}(t, {\bf x}).
\eq
 Abandoning  the Lorentz symmetry   gives rise  to   a proliferation of independently  coupling constants \cite{BPS,KP},
which  could potentially limit   the prediction powers of the theory. To reduce the number of these constants, Horava
imposed two conditions, {\em the projectability   and detailed balance } \cite{Horava}. The
former assumes that the lapse function $N$ in the  Arnowitt-Deser-Misner   decompositions  \cite{ADM} is a function of
$t$ only,
\bq
\lb{1.6}
N = N(t),
\eq
while the latter assumes that  gravitational potential ${\cal{L}}_{V}$ can be obtained from a superpotential $W_{g}$
via the relations,
\bq
\lb{1.7}
{\cal{L}}_{(V, D)} = E_{ij}{\cal{G}}^{ijkl}E_{kl},\;\;\;
E^{ij} = \frac{1}{\sqrt{g}}\frac{\delta{W}_{g}}{\delta{g}_{ij}},
\eq
where  ${\cal{G}}^{ijkl}$ denotes the generalized De-Witt metric, defined as
${\cal{G}}^{ijkl} = \frac{1}{2} \big(g^{ik}g^{jl} + g^{il}g^{jk}\big) - \lambda g^{ij}g^{kl}$,
and $\lambda$ is a coupling constant.

 However, with the detailed balance  condition, 
the Newtonian  limit does not exist \cite{LMP}, and  a scalar field in the UV is not stable  \cite{CalA}. Thus, it is
generally believed that this condition should be abandoned  \cite{KK}.  But,  due to    several remarkable features \cite{Hreview},
Borzou, Lin, and Wang  recently studied it in detail, and found that   the scalar field  can be stabilized, if the detailed balance  condition is
allowed to be softly broken \cite{BLW}. With such a breaking, all  the other related problems
found so far also can be resolved. For detail, we refer readers to \cite{BLW}.

On the other hand, with the projectability condition, the number of independent coupling  constants can be significantly reduced. In fact, together with the
assumptions of the parity and time-reflection symmetry, it can be reduced from more than 70 to 11 \cite{SVW} (See also \cite{KKa}). But, the Minkowski spacetime
now becomes unstable \cite{SVW,BS,WM}, although the de Sitter spacetime is \cite{HWW,WWa}. In addition, such a theory also faces
 the strong coupling problem \cite{SC,WWa} \footnote{In the literature, the ghost problem was often mentioned \cite{reviews,Mukc}. But, by restricting the
coupling constant $\lambda$ to the regions $\lambda \ge 1$ or $\lambda < 1/3$, this problem is solved (at least in the classical level) \cite{Horava,SVW,BS,WM}. In addtion, when $\lambda
\in(1/3, 1)$, the instability problem disappears. Therefore, one of these two problems can  be always avoided by properly choosing $\lambda$.
In this paper, we choose  $\lambda \ge 1$, so the  ghost problem does not exist.}. It should be noted that both of these two problems are closely related to the existence of a spin-0
graviton \cite{reviews,Mukc}, because of the   foliation-preserving diffeomorphisms (\ref{1.4}) \footnote{Since  the   foliation-preserving diffeomorphisms (\ref{1.4})  is also
assumed in  the version  without the projectability condition \cite{BPS}, the spin-0 graviton exists there too.}. Another problem related to  the presence of this spin-0 graviton is
the difference of its speed from that of   the spin-2 graviton.  Since   they are not related by any symmetry, it  poses
a great challenge for any attempt to restore Lorentz symmetry  at low energies where it has been  well tested experimentally.  In particular, one needs a mechanism
to ensure that in those energy scales all species  of matter and gravity  have the same effective speed and light cones.

To overcome these problems, so far three main different approaches have been taken. The first one is
to provoke the Vainshtein mechanism, initially found in massive gravity \cite{Vain}. In particular, Mukohyama studied spherically symmetric static
spacetimes \cite{Mukc}, and showed that the spin-0 gravitons decouple after nonlinear effects are taken into account. Similar considerations in cosmology were given
in \cite{Izumi:2011eh,GMW} (See also \cite{WWa}), where a fully nonlinear analysis of superhorizon cosmological perturbations was carried out,
by adopting the so-called gradient expansion method \cite{SBLMS}. It was found that   the relativistic  limit of the Horava-Lifshitz (HL)  theory  is continuous,
and general relativity (GR) is recovered at least in two different cases: (a) when only  the ``dark matter as an integration constant'' is present \cite{Izumi:2011eh};
and (b) when a scalar field and    the ``dark matter as an integration constant'' are present \cite{GMW}.

 Another very attractive and completely different approach is to eliminate the spin-0 gravitons and meanwhile fix
$\lambda$ to its relativistic value,  $\lambda_{GR} = 1$. This was   done  recently by Horava and Melby-Thompson (HMT)
\cite{HMT}. HMT first noticed that in the linearized theory  a $U(1)$ symmetry exists only  in  the case $\lambda = 1$
\cite{Horava}. Thus, to fix $\lambda$, one may extend the foliation-preserving diffeomorphism symmetry (\ref{1.4})  to
\bq
\lb{symmetry}
 U(1) \ltimes {\mbox{Diff}}(M, \; {\cal{F}}).
 \eq
To  lift such a   symmetry to the full nonlinear theory, HMT found that it is necessary to introduce a
scalar field - the Newtonian prepotential, in addition to the $U(1)$ gauge field. Once this was done,
 HMT  showed   that  the spin-0 graviton is eliminated \cite{HMT}. This was further confirmed in \cite{WW}. 
Then, the instability and strong coupling problems  of the spin-0 gravitons are out of question.  In addition, since $\lambda$ was fixed to $1$,
these   problems in the nongravitational sectors are also resolved, as all of them are related to the  fact that $\lambda \not= 1$ \cite{HW}.

However,  da Silva soon found that the introduction of the Newtonian
prepotential is so strong that actions with $\lambda \not=1$ also
have the   extended symmetry  (\ref{symmetry}) \cite{Silva}. The
spin-0 gravitons  are eliminated even with any given $\lambda$
\cite{Silva,HW,Kluson2},  so that the strong coupling problem does
not exist any longer in the pure gravitational sector. However,   it
still exists when   matter is present. Indeed, in   \cite{HW} it was
shown that,  for  processes with energy higher than
$\Lambda_{\omega} [\equiv |\lambda - 1|^{5/2}M_{pl}]$, the theory
becomes strong
 coupling \cite{HW}.  Together with Lin, three of the present authors  \cite{LWWZ} showed  that this problem can be resolved by introducing
a new energy scale $M_{*}$  \cite{BPSc}, so that $M_{*} < \Lambda_{\omega}$, where $M_{*}$ denotes the suppression energy scale
of  high-order derivative terms of the theory.

Note that the above two approaches assume the projectability condition (\ref{1.6}). The third approach is to abandon this condition, by including the vector field \cite{BPS}
\bq
\lb{vector}
a_{i} = \partial_{i}\ln(N),
\eq
into the action \footnote{ It should be noted that  the  violation of  the projectability condition often leads to the inconsistency
problem \cite{LP}. However, as shown in \cite{Kluson}, this is not the case  in the setup of \cite{BPS}.}.   Although it also solves the instability and strong coupling problems,    the presence of this vector field
$a_{i}$ gives rise  to a proliferation of independent  coupling constants \cite{KP}, as mentioned above. When applying the theory to cosmology and astrophysics, this  potentially limits its predictive powers.
In addition,  the problem of different speeds in the gravitational sector still exists, because the spin-0 graviton still exists in this setup, and its speed depends on the coupling constants $\lambda$ and
$\beta_0$ \cite{BPS,CHZ,KUY}, while the problem of the spin-2 graviton is independent of them, where $\beta_0$ is defined in Eq.(\ref{potential}) given below.

Recently, we proposed an extended version of HL gravity without the projectibility condition (\ref{1.6}) but with the enlarged symmetry (\ref{symmetry}) \cite{ZWWS}, with
the purposes: (i) Reduce significantly the number of the independent coupling constants usually presented in the version of the HL theory without the projectability condition,
by imposing the detailed balance condition. However, in order for the theory to  be both UV complete and  IR  healthy, we allowed the detailed balance
condition to be broken softly by adding all the low dimensional relevant terms. 
(ii) Eliminate the spin-0 gravitons even in the case without the projectability condition by implementing  the enlarged symmetry (\ref{symmetry})
\footnote{Note that the $U(1)$ symmetry in the case without projectability condition was also considered in the so-called $F(R)$ Horava-Lifshitz gravity \cite{Klusonb}.},
so that all the problems related to them disappear, including the instability,  strong coupling and different speeds in the pure gravitational sector.

In this paper, we shall first provide a systematical study of this extended version of the HL gravity regarding to the above mentioned problems in  the   gravitational  as well as matter sectors, and then apply it to cosmology.
In particular, the paper is organized as follows. In Sec II, we construct the gravitational potential  by imposing the detailed balance condition softly breaking. In Sec III, we extend the foliation-preserving diffeomorphism symmetry of HL gravity to include a local $U(1)$ symmetry, and with this enlarged symmetry, in Sec IV, we show that the spin-0 gravitons are indeed eliminated. In Sec V, we consider the coupling of the theory with a scalar field, and show that the scalar field is stable in both of the UV and IR. In addition, the strong coupling problem does not exist, because of the presence of the sixth-order spatial derivative terms of the scalar field, as
long as their suppressed energy scale $M_{*}$ is lower than the would-be strong coupling energy scale $\Lambda_{\omega}$. In Sec VI we study cosmological models, and show that the FRW universe is necessarily flat
in such a setup, while   in Sec VII, we investigate the linear scalar, vector and tensor perturbations of the flat FRW universe, and present the general linearized field equations for each kind of the perturbations.  Finally, in Sec VIII we present our main conclusions.

\section{Potential with Detailed Balance Condition Softly Breaking}

\renewcommand{\theequation}{2.\arabic{equation}} \setcounter{equation}{0}

 To understand the consequence of the breaking of the projectability condition (\ref{1.6}), let us start with counting the independent terms order by order. We first write
 the metric in the  Arnowitt-Deser-Misner   form  \cite{ADM},
 \bqn
 \lb{1.2}
ds^{2} &=& - N^{2}c^{2}dt^{2} + g_{ij}\left(dx^{i} + N^{i}dt\right)
     \left(dx^{j} + N^{j}dt\right), \nb\\
     & & ~~~~~~~~~~~~~~~~~~~~~~~~~~~~~~  (i, \; j = 1, 2, 3).~~~
 \eqn
 Under the rescaling (\ref{1.1})
 with   $z  = 3$,     $N, \; N^{i}$ and $g_{ij}$ scale,
 respectively,  as,
 \bq
 \lb{1.3}
  N \rightarrow  N ,\;  N^{i}
\rightarrow b^{-2} N^{i},\; g_{ij} \rightarrow g_{ij}.
 \eq
 Under  the  foliation-preserving diffeomorphisms (\ref{1.4}), they transform as 
\bqn
\lb{1.5}
\delta{g}_{ij} &=& \nabla_{i}\zeta_{j} + \nabla_{j}\zeta_{i} + f\dot{g}_{ij},\nb\\
\delta{N}_{i} &=& N_{k}\nabla_{i}\zeta^{k} + \zeta^{k}\nabla_{k}N_{i}  + g_{ik}\dot{\zeta}^{k}
+ \dot{N}_{i}f + N_{i}\dot{f}, \nb\\
\delta{N} &=& \zeta^{k}\nabla_{k}N + \dot{N}f + N\dot{f},
\eqn
where $\dot{f} \equiv df/dt$,  $\nabla_{i}$ denotes the covariant
derivative with respect to the 3-metric $g_{ij}$ and $N_{i} = g_{ik}N^{k}$.

 Assuming that the engineering dimensions of space and time are \cite{SVW},
 \bq
 \lb{2.1}
[dx] = [k]^{-1}, \;\;\;  [dt] = [k]^{-3},
 \eq
 we find that
 \bqn
 \lb{2.2}
&&  [N^{i}] = [c] = \frac{[dx]}{[dt]} = [k]^{2},\;\;\;  [g_{ij}] = [N] = [1],\nb\\
&& [K_{ij}] = [k]^{3},\;\;\;  [\Gamma^{i}_{jk}] = [k],\;\;\; [R^{i}_{jkl}] = [k]^{2}.
 \eqn
Then, to each order of $[k]$, we have the following independent terms that are all scalars under the transformations of the
foliation-preserving
diffeomorphisms (\ref{1.4}) \cite{SVW,BPS,KP},
\bqn
\lb{terms}
&& [k]^{6}: K_{ij}K^{ij},\; K^{2},\; R^{3}, \;   RR_{ij}R^{ij} , \; R^{i}_{j}R^{j}_{k}R^{k}_{i},\;  \left(\nabla{R}\right)^{2}, \nb\\
&& ~~~~~~~~ \left(\nabla_{i}R_{jk}\right) \left(\nabla^{i}R^{jk}\right),\;  \left(a_{i}a^{i}\right)^{2}R , \;
  \left(a_{i}a^{i}\right)\left(a_{i}a_{j}R^{ij}\right),\nb\\
&& ~~~~~~~~ \left(a_{i} a^{i}\right)^{3}, a^{i}\Delta^{2}a_{i},  \left(a^{i}_{\;\;i}\right) \Delta{R},  ...,\nb\\
&& [k]^{5}: \;  K_{ij}R^{ij},\; \epsilon^{ijk}R_{il}\nabla_{j}R^{l}_{k},\;  \epsilon^{ijk}a_{i}a_{l}\nabla_{j}R^{l}_{k}, \nb\\
&& ~~~~~~~~  a_{i}a_{j} K^{ij}, \; K^{ij}a_{ij},\;    \left(a^{i}_{\;\;i}\right)K, \nb\\
&& [k]^{4}: \; R^{2},\; R_{ij}R^{ij},\; \left(a_{i}a^{i}\right)^{2},  \; \left(a^{i}_{\;\;i}\right)^{2},\;  \left(a_{i}a^{i}\right)a^{j}_{\;\;j},\nb\\
&&  ~~~~~~~~  a^{ij}a_{ij},\;  \left(a_{i}a^{i}\right)R,\;  a_{i}a_{j}R^{ij},\;  Ra^{i}_{\;\;i},   \nb\\
&& [k]^{3}: \; \omega_{3}(\Gamma),\nb\\
&& [k]^{2}: \; R,\;  a_{i}a^{i},  \nb\\
&& [k]^{1}: \; {\mbox{None}},\nb\\
&& [k]^{0}: \; \gamma_{0},
\eqn
where  $\omega_{3}(\Gamma)$ denotes the gravitational Chern-Simons term, $\gamma_{0}$ is a dimensionless constant,
$\epsilon^{ijk}  \equiv {e^{ijk}}/{\sqrt{g}}, (e^{123} = 1)$, $\Delta = g^{ij}\nabla_{i}\nabla_{j}$, and
\bqn
\lb{2.3}
 K_{ij} &=& \frac{1}{2N}\left(-\dot{g}_{ij} + \nabla_{i}N_{j} + \nabla_{j}N_{i}\right),\nb\\
 a_{i_{1}i_{2}...i_{n}}  &=& \nabla_{i_{1}} \nabla_{i_{2}}... \nabla_{i_{n}}\ln(N),\nb\\
\omega_{3}(\Gamma) &=& {\mbox{Tr}} \Big(\Gamma \wedge d\Gamma + \frac{2}{3}\Gamma\wedge \Gamma\wedge \Gamma\Big)\nb\\
&=& \frac{e^{ijk}}{\sqrt{g}}\Big(\Gamma^{m}_{jl}\partial_{j}\Gamma^{l}_{km} + \frac{2}{3}\Gamma^{n}_{il}\Gamma^{l}_{jm} \Gamma^{m}_{kn}\Big).
\eqn
In writing Eq.(\ref{terms}), we had not written down all the sixth  order terms, as they are numerous \cite{BPS,KP}.
Then, the general action of the gravitational part will be given by
\bq
\lb{action}
\hat{S}_{g} = \zeta^{2}\int{dtd^{3}x N\sqrt{g}\Big({\cal{L}}_{K}  - {\cal{L}}_{V}\Big)},
\eq
where the kinetic part ${\cal{L}}_{K}$ is the linear combination of the first two sixth order derivative terms, 
\bq
\lb{2.4a}
{\cal{L}}_{K} = g_{T}\left(K_{ij}K^{ij}- \lambda K^{2}\right).
\eq
Note that the coupling constant $g_{T}$ can be absorbed into $\zeta^{2}$. So, without loss of generality, we can set it to one,
$g_{T} = 1$. The potential part  ${\cal{L}}_{V}$ is the linear combination of all the other terms of Eq.(\ref{terms}),
which are more than 70 terms, and could potentially weaken  the  prediction powers of the theory.

In the following, we look for conditions to reduce the number of the
independent terms. First, since those with odd number of derivatives
violate the spatial parity and time-reversal symmetry, they can be
easily eliminated by imposing the parity conservation and
time-reversal symmetry. To reduce  the number of   the sixth order
derivative terms, following Horava we impose the ``generalized"
detailed balance condition,
\bq
\lb{2.4}
\hat{\cal{L}}_{(V, D)} = {\cal{L}}_{(V,D)} - g_{ij}A^{i}A^{j},
\eq
where   $A^{i}$ is defined by the superpotential $W_{a}$,
\bq
\lb{2.5}
A^{i} = \frac{1}{\sqrt{g}}\frac{\delta{W_{a}}}{\delta{a_{i}}},
\eq
with
\bq
\lb{2.5a}
W_{a} = \frac12 \int{d^{3}x \sqrt{g} a^{i}\left(\sum_{n=
0}^{n=1}{\mathfrak{b}_{n}\Delta^{n}{a_{i}}}\right)},
\eq
where $\mathfrak{b}_{n}$ are arbitrary  constants. Note that   the term of $a_{i}\Delta^{1/2}a^{i}$  in principle can  be included into $W_{a}$, which will give rise to fractional calculus, a
branch of mathematics that has been well developed \cite{FC}. But this gives rise to fifth order derivative terms, and  we shall discard  these  terms.
 Inserting Eq.(\ref{2.5}) into Eq.(\ref{2.4}), we find that its second term is  involved only with $a_{i}$, and the
corresponding  action  takes the form,
\bq
\lb{2.6}
S_{a} = \int{dtd^{3}N\sqrt{g}{\Big[\beta_0 \left(a_{i}a^{i}\right) + {\eta_{1}} a_{i}\Delta{a^{i}} + {\eta_{2}} \left(\Delta{a^{i}}\right)^{2}\Big]}},
\eq
where
\bq
\lb{2.7}
\beta_0 \equiv \mathfrak{b}_{0}^{2},\;\;\; \eta_1 \equiv 2\mathfrak{b}_{0}\mathfrak{b}_{1},\;\;\;
\eta_2 \equiv \mathfrak{ b}_{1}^{2}.
\eq

The superpotential  $W_{g}$ appearing in Eq.(\ref{1.7}) is   given by \cite{Horava}
\bq
\lb{1.8}
W_{g}  = \int_{\Sigma}{\left(\frac{1}{w^{2}}\omega_{3}(\Gamma) + \mu \big(R - 2 \Lambda_{W}\big)\right)},
\eq
where $\omega_{3}(\Gamma)$ is defined in Eq.(\ref{2.3}), and $R$  ($R_{ij}$) is the Ricci scalar (tensor)  built out of $g_{ij}$. Inserting Eq.(\ref{1.8}) into Eq.(\ref{1.7}), we find that
\bqn
\lb{1.11}
{\cal{L}}_{(V,D)}   &=& \zeta^{2}\gamma_{0}  + \gamma_{1} R + \frac{1}{\zeta^{2}}
\Big(\gamma_{2}R^{2} +  \gamma_{3}  R_{ij}R^{ij}\Big)\nb\\
& &  +  \frac{\gamma_{4}}{\zeta^{3}} \epsilon^{ijk}R_{il}\nabla_{j}R^{l}_{k} +  \frac{\gamma_{5}}{\zeta^{4}} C_{ij}C^{ij}, ~~~
 \eqn
where  $\gamma_{n}$ are dimensionless constants, given explicitly in terms of the five independent coupling constants
$\zeta, \; w,\; \mu,\; \Lambda_{W}$, and $\lambda$ in \cite{Horava}. $C_{ij}$ denotes the Cotton tensor, defined by
\bq
\lb{1.12}
C^{ij} = \frac{ {{e}}^{ikl}}{\sqrt{g}} \nabla_{k}\Big(R^{j}_{l} - \frac{1}{4}R\delta^{j}_{l}\Big).
\eq
Using the Bianchi identities and the definition of the Riemann tensor, one can show that $C_{ij}C^{ij}$ can be written in terms of the
five independent sixth-order derivative terms in the form
\bqn
\lb{1.13}
C_{ij}C^{ij} 
&=& \frac{1}{2}R^{3} - \frac{5}{2}RR_{ij}R^{ij} + 3 R^{i}_{j}R^{j}_{k}R^{k}_{i}  +\frac{3}{8}R\Delta R\nb\\
& &  +
\left(\nabla_{i}R_{jk}\right) \left(\nabla^{i}R^{jk}\right) +   \nabla_{k} G^{k},
\eqn
where
\lb{1.14}
\bqn
G^{k}=\frac{1}{2} R^{jk} \nabla_j R - R_{ij} \nabla^j R^{ik}-\frac{3}{8}R\nabla^k R.
\eqn
When integrated, with the projectability condition (\ref{1.6}),  $\nabla_{k}G^{k}$  becomes a  boundary term and can be discarded. However,   in the case without this condition,
this is no longer true, since now we have $N = N(t, x)$ and
\bq
\lb{1.14a}
\int_{M}{dt d^{3}x N\sqrt{g} \nabla_{k}G^{k}} = 
 - \int_{M}{dt d^{3}x N\sqrt{g}G^{k} a_{k}},
\eq
which in general is not zero.

As mentioned previously, in order for the theory to have a healthy IR limit,     the detailed condition needs to be broken softly by adding all the lower
(than six) dimensional relevant terms presented in Eq.(\ref{terms}), so that finally the potential  is given by \cite{ZWWS}
\bqn
\lb{potential}
{\cal{L}}_{V} &=&  \gamma_{0}\zeta^{2}  -  \Big(\beta_0  a_{i}a^{i}- \gamma_1R\Big)
+ \frac{1}{\zeta^{2}} \Big(\gamma_{2}R^{2} +  \gamma_{3}  R_{ij}R^{ij}\Big)\nb\\
& & + \frac{1}{\zeta^{2}}\Bigg[\beta_{1} \left(a_{i}a^{i}\right)^{2} + \beta_{2} \left(a^{i}_{\;\;i}\right)^{2}
+ \beta_{3} \left(a_{i}a^{i}\right)a^{j}_{\;\;j} \nb\\
& & + \beta_{4} a^{ij}a_{ij} + \beta_{5}
\left(a_{i}a^{i}\right)R + \beta_{6} a_{i}a_{j}R^{ij} + \beta_{7} Ra^{i}_{\;\;i}\Bigg]\nb\\
& &   
 +  \frac{1}{\zeta^{4}}\Bigg[\gamma_{5}C_{ij}C^{ij}  + \beta_{8} \left(\Delta{a^{i}}\right)^{2}\Bigg],
\eqn
where $ \beta_{8} \equiv - \eta_2 \zeta^{4}$. All the coefficients, $ \beta_{n}$ and $\gamma_{n}$, are
dimensionless and arbitrary, except for the ones of the sixth-order derivative terms,  $\gamma_{5}$ and $\beta_{8}$, which
must be
\bq
\lb{2.8a}
 \gamma_{5} > 0, \;\;\; \beta_{8} <  0,
\eq
as can be seen from Eqs.(\ref{1.8}) and (\ref{2.7}). To be consistent with observations in the IR, we must set
\bq
\lb{2.8}
\zeta^{2} = \frac{1}{16\pi G},\;\;\;  \gamma_{1} = -1,
\eq
where $G$ denotes the Newtonian constant, and
\bq
\lb{2.8b}
\Lambda \equiv \frac{1}{2} \zeta^{2}\gamma_{0},
\eq
is the cosmological constant.

It can be shown that for quadratic action of the scalar perturbations in the Minkowski background  the  sixth-order spatial derivative
terms of the potential (\ref{potential}) are absent. As a result, the gravitational sector is still strong coupling, and cannot be solved    by the mechanism proposed in \cite{BPSc}. To solve this problem, one
way is to eliminate the spin-0 gravitons, as HMT did in the case with the projectability condition. In the next section, we will show explicitly that this is possible by enlarging the
$ {\mbox{Diff}}(M, \; {\cal{F}})$ symmetry (\ref{1.4}) to  the one  $U(1) \ltimes {\mbox{Diff}}(M, \; {\cal{F}})$ (\ref{symmetry}),  even in the case without the  projectability condition.

\section{$U(1) \ltimes {\mbox{Diff}}(M, {\cal{F}})$ Symmetry and Field Equations}

\renewcommand{\theequation}{3.\arabic{equation}} \setcounter{equation}{0}

In order to   eliminate the spin-0 gravitons, let us first consider the $U(1)$ gauge
transformations \cite{HMT},
\bqn
\lb{gauge}
\delta_\alpha N_i=N\nabla_i\alpha,~~~~~\delta_\alpha g_{ij}=0=\delta_\alpha N,
\eqn
where $\alpha$ denotes the $U(1)$ generator. Under the above transformations, the variation of the HL action (\ref{action}) is given by
\bqn
\lb{actiontrans}
\delta\hat{S}_g&=&\zeta^2\int dtd^3x \sqrt{g}(\dot{\alpha}-N^i\nabla_i\alpha)R,\nb\\
&& +2\zeta^2\int dtd^3x \sqrt{g}N\alpha G^{ij}K_{ij},\nb\\
&& + 2\zeta^2\int dtd^3x \sqrt{g}N\hat{{\cal{G}}}^{ijlk}K_{ij} a_{(l}\nabla_{k)}\alpha,\nb\\
&& +2(1-\lambda)\zeta^2\int dtd^3x \sqrt{g}NK(\nabla^2\alpha+a^k\nabla_k\alpha),\nb\\
\eqn
where $f_{(ij)}=( f_{ij}+f_{ji})/2$,
$\hat{{\cal{G}}}^{ijlk}=g^{il}g^{jk}- g^{ij}g^{lk}$, and  $G^{ij} = R^{ij} - g^{ij}R/2$.
In order for the theory  to  have the $U(1)$ symmetry, one can introduce a  $U(1)$ gauge field $A$, which transforms as
\bqn
\lb{gaugeAA}
\delta_\alpha A=\dot{\alpha}-N^i\nabla_i\alpha.
\eqn
Then, by adding the new coupling term
\bqn
\lb{actionA}
S_A&=& \zeta^2\int dtd^3xN\sqrt{g}{\cal{L}}_{A} \nb\\
&=&  \zeta^2\int dtd^3x\sqrt{g}A(2\Lambda_{g} -R),
\eqn
to $\hat{S}_{g}$, one finds that its variation (for $\Lambda_{g} = 0$) with respect to $\alpha$ exactly cancels the first term given in Eq.(\ref{actiontrans}).
To repair the rest,   we introduce  the Newtonian prepotential $\varphi$, which transforms as
\bq
\lb{eqb}
\delta_\alpha\varphi= - \alpha.
\eq
Then, it can be shown that under Eq.(\ref{gauge})  the variation of  the term
\bqn
\lb{svarphi1}
S_{(\varphi,1)}
&=&  \zeta^2\int dtd^3x\sqrt{g}N\varphi {\cal{G}}^{ij}\Big[2K_{ij} + a_{(i}\nabla_{j)}\varphi \nb\\
&& ~~~~~~~ + \nabla_{i}\nabla_{j}\varphi\Big],
\eqn
 exactly cancels the second term in Eq.(\ref{actiontrans}) as well as the term $2A\Lambda_{g}$ in (\ref{actionA}),
 where
\bqn
{{\cal{G}}}^{ij}\equiv R^{ij}-\frac{1}{2}Rg^{ij}+\Lambda_g g^{ij}.
\eqn
The third and fourth  terms in (\ref{actiontrans}) can be canceled, respectively,  by
 \bqn
\lb{svarphi2}
S_{(\varphi,2)}
&=&  \frac{\zeta^2}3\int dtd^3x\sqrt{g}N\hat{\cal{G}}^{ijkl}\Big[6K_{ij}a_{(k}\nabla_{l)}\varphi \nb\\
&& + 4\left(\nabla_{i}\nabla_{j}\right) a_{(k}\nabla_{l)}\varphi + 5 a_{(i}\nabla_{j)}\varphi a_{(k}\nabla_{l)}\varphi \nb\\
&&  + 2\nabla_{(i}\varphi a_{j)(k}\nabla_{l)}\varphi\Big],
\eqn
and
 \bqn
\lb{svarphi3}
S_{(\varphi,3)}
&=&   (1-\lambda)  \zeta^2\int dtd^3x\sqrt{g}N\Bigg\{\left(a^{k}\nabla_{k}\varphi + \nabla^{2}\varphi\right)\nb\\
&& \times \Big[2K +  \left(a^{k}\nabla_{k}\varphi + \nabla^{2}\varphi\right)\Big]\Bigg\}.
\eqn
Hence, the action
\bq
\lb{actions}
S_{g} = \hat{S}_{g} + S_{A} + S_{\varphi},
\eq
 is invariant under the $U(1) \ltimes {\mbox{Diff}}(M, {\cal{F}})$ symmetry (\ref{symmetry}),
 where
\bqn
\lb{action nu}
S_\varphi&=& \sum_{n=1}^{3}{S_{(\varphi,n)}} \equiv \zeta^2\int dtd^3xN\sqrt{g}{\cal{L}}_{\varphi},
\eqn
with
\bqn
\lb{action nuA}
{\cal{L}}_{\varphi} &=&  \varphi{\cal{G}}^{ij}\big(2K_{ij}+\nabla_i\nabla_j\varphi+a_i\nabla_j\varphi\big)\nb\\
& & +(1-\lambda)\Big[\big(\nabla^2\varphi+a_i\nabla^i\varphi\big)^2 
+2\big(\nabla^2\varphi+a_i\nabla^i\varphi\big)K\Big]\nb\\
& & +\frac{1}{3}\hat{\cal G}^{ijlk}\Big[4\left(\nabla_{i}\nabla_{j}\varphi\right) a_{(k}\nabla_{l)}\varphi + 5 \left(a_{(i}\nabla_{j)}\varphi\right) a_{(k}\nabla_{l)}\varphi\nb\\
&&
+ 2 \left(\nabla_{(i}\varphi\right)a_{j)(k}\nabla_{l)}\varphi
+ 6K_{ij} a_{(l}\nabla_{k)}\varphi
\Big].
\eqn

When coupling to the matter ${\cal{L}}_M$, the total action of the theory takes the form
\bqn\label{total action}
S&=&\zeta^2 \int dtd^3x \sqrt{g} N \Bigg({\cal{L}}_{K}-{\cal{L}}_{V} + {\cal{L}}_{A}
+ {\cal{L}}_{\varphi}\nb\\
&& ~~~~~~~~~~~~~~~~~~~~~~~~~  + \frac{1}{\zeta^{2}} {\cal{L}}_M\Bigg).
\eqn
Then, the variations of $S$ with respect to $N$ and $N^{i}$ give rise to the Hamiltonian and momentum constraints,
\bqn \label{hami}
 {\cal{L}}_K + {\cal{L}}_V^R + F_V-F_\varphi-F_\lambda= 8 \pi G  J^t,\;\;
\eqn
\bqn \label{mom}
&&\nabla_j \bigg\{\pi^{ij} - \varphi {\cal{G}}^{ij} - \hat{{\cal{G}}}^{ijkl} a_l \nabla_k \varphi\nb\\
&&-(1-\lambda)g^{ij}\big(\nabla^2\varphi+a_k\nabla^k\varphi\big)\bigg\}=8\pi G J^i,
\eqn
where
\bqn
{\cal{L}}_V^R&=&\gamma_0 \zeta^2-R+\frac{\gamma_2 R^2+\gamma_3 R_{ij}R^{ij}}{\zeta^2}+\frac{\gamma_5}{\zeta^4} C_{ij}C^{ij},\nb\\
J^i &=& -N \frac{\delta {\cal{L}}_M}{\delta N_i},\;\;
J^t = 2 \frac{\delta (N {\cal{L}}_M)}{\delta N},\nb\\
\pi^{ij}&=&-K^{ij}+\lambda K g^{ij},
\eqn
and $F_V,\;F_\varphi$, and $F_\lambda$ are given by Eqs.(\ref{a1})-(\ref{a3}) in Appendix A. Note that we have separated ${\cal{L}}_V$ into two parts ${\cal{L}}_V^R$ and ${\cal{L}}_V^a$. Variations of $S$ with respect to $\varphi$ and $A$ yield, respectively,
\bqn \label{phi}
&& \frac{1}{2} {\cal{G}}^{ij} ( 2K_{ij} + \nabla_i\nabla_j\varphi  +a_{(i}\nabla_{j)}\varphi)\nb\\
&& + \frac{1}{2N} \bigg\{ {\cal{G}}^{ij} \nabla_j\nabla_i(N \varphi) - {\cal{G}}^{ij} \nabla_j ( N \varphi a_i)\bigg\}\nb\\
&& - \frac{1}{N} \hat{{\cal{G}}}^{ijkl} \bigg \{ \nabla_{(k} ( a_{l)} N K_{ij}) + \frac{2}{3} \nabla_{(k} (a_{l)} N \nabla_i \nabla_j \varphi)\nb\\
&& - \frac{2}{3} \nabla_{(j} \nabla_{i)} (N a_{(l} \nabla_{k)} \varphi) + \frac{5}{3} \nabla_j (N a_i a_k \nabla_l \varphi)\nb\\
&& + \frac{2}{3} \nabla_j (N a_{ik} \nabla_l \varphi)\bigg\} \nb\\
&& + \frac{1-\lambda}{N} \bigg\{\nabla^2  \left.[N (\nabla^2 \varphi + a_k \nabla^k \varphi)\right.] \nb\\
&& - \nabla^i [N(\nabla^2 \varphi + a_k \nabla^k \varphi) a_i] \nb\\
&&+ \nabla^2 (N K) - \nabla^i ( N K a_i)\bigg \}
 = 8 \pi G J_\varphi,
\eqn
and
\bqn \label{A}
R-2 \Lambda_g = 8 \pi G J_A,
\eqn
where
\bqn
J_\varphi = -\frac{\delta {\cal{L}}_M}{\delta \varphi},\;\;\;
 J_A= 2 \frac{\delta ( N {\cal{L}}_M)}{\delta A}.
\eqn
On the other hand, the variation of $S$ with respect to $g_{ij}$ yields the dynamical equations,
\bqn \label{dyn}
\frac{1}{\sqrt{g}N} \frac{\partial}{\partial t}\left(\sqrt{g} \pi^{ij}\right)+2(K^{ik}K^j_k-\lambda K K^{ij})\nb\\
+\frac{1}{N}\nabla_k (\pi^{ik}N^j+\pi^{kj}N^i-\pi^{ij}N^k)\nb\\
-F^{ij}-F^{ij}_a-F^{ij}_\varphi-\frac{1}{2}g^{ij}{\cal{L}}_K-\frac{1}{2}g^{ij}{\cal{L}}_A\nb\\
-\frac{1}{N}(AR^{ij}+g^{ij}\nabla^2A-\nabla^j\nabla^iA)
=8\pi G \tau^{ij},\;\;\;\;\;\;
\eqn
where
\bqn
\lb{tauij}
\tau^{ij}&=&\frac{2}{\sqrt{g}N} \frac{\delta(\sqrt{g}N{\cal{L}}_M)}{\delta g_{ij}}, \nb\\
F^{ij}&=&\frac{1}{\sqrt{g}N}\frac{\delta (-\sqrt{g}N {\cal{L}}_V^R)}{\delta g_{ij}}\nb\\
&=& \sum_{s=0}\hat{\gamma}_s\zeta^{n_s}(F_s)^{ij},\;\;\;\\
F^{ij}_a&=&\frac{1}{\sqrt{g}N}\frac{\delta (-\sqrt{g}N {\cal{L}}_V^a)}{\delta g_{ij}}\nb\\
& =& \sum_{s=0}\beta_s\zeta^{m_s}(F_s^a)^{ij},\\
F^{ij}_\varphi&=&\frac{1}{\sqrt{g}N}\frac{\delta (\sqrt{g}N {\cal{L}}_\varphi)}{\delta g_{ij}}\nb\\
&=& \sum_{s=0}\mu_s(F_s^\varphi)^{ij}.
\eqn
The expressions of $F_s$, $F_s^a$ and $F_s^\varphi$ can be found in Appendix (\ref{a4})-(\ref{a6}), and
\bqn
\hat{\gamma}_s &=& \left(\gamma_0, \gamma_1, \gamma_2, \gamma_3, \frac{1}{2}\gamma_5, -\frac{5}{2}\gamma_5, 3\gamma_5, \frac{3}{8}\gamma_5, \gamma_5, \frac{1}{2}\gamma_5\right), \nb\\
n_s &=& (2, 0, -2, -2, -4, -4, -4, -4, -4, -4),\nb\\
m_s&=& (0, -2,-2,-2, -2, -2, -2, -2, -4 ), \nb\\
\mu_s &=& \left(2, 1, 1, 2, \frac{4}{3}, \frac{5}{3}, \frac{2}{3}, 1-\lambda, 2-2 \lambda\right).
\eqn

In addition, the matter components $(J^t, J^i, J_\varphi, J_A, \tau^{ij})$ satisfy the conservation laws of energy and momentum,
\bqn
\label{energy conservation}
\int d^3x \sqrt{g} N \bigg[\dot{g}_{ij}\tau^{ij}-\frac{1}{\sqrt{g}}\partial_t (\sqrt{g} J^t)+\frac{2 N_i}{\sqrt{g} N}\partial_t (\sqrt{g} J^i)\nb\\
-\frac{A}{\sqrt{g} N}\partial_t (\sqrt{g} J_A)-2\dot{\varphi}J_\varphi\bigg]=0,\;\;\;\;\;\;\;\;\\
\label{mom conservation}
\frac{1}{N}\nabla^i(N\tau_{ik})-\frac{1}{\sqrt{g} N} \partial_t (\sqrt{g} J_k)-\frac{J_A}{2N}\nabla_k A-\frac{J^t}{2N}\nabla_kN\nb\\
-\frac{N_k}{N} \nabla_i J^i-\frac{J_i}{N}(\nabla_i N_k-\nabla_k N_i)+J_\varphi \nabla_k\varphi=0.\;\;\;\;\;\;\;\;
\eqn

\section{Elimination of Spin-0 Gravitons}
\renewcommand{\theequation}{4.\arabic{equation}} \setcounter{equation}{0}

In this section, we show that the spin-0 gravitons are indeed eliminated for the theory described by the action (\ref{actions}).
To this goal, we consider the scalar perturbations of the Minkowski background,
\bqn
\lb{Minkp}
N=1+\phi, ~N_i=\partial_i B,\nb\\
 g_{ij}=(1-2\psi)\delta_{ij}+ 2E_{,ij},\nb\\
A=\delta A,~~\varphi=\delta \varphi.
\eqn
Using the gauge freedom, without loss of generality, one can choose the gauge
\bqn\lb{gaugeA}
E=0, \;\;\; \varphi=0.
\eqn
Then,  after simple but tedious calculations, to second order we find that 
\bqn
\lb{second action}
S^{(2)}&=&\zeta^2\int dtd^3x\bigg\{(1-3\lambda)(3\dot{\psi}^2+2\dot{\psi}\partial^2B)\nb\\
& &+(1-\lambda)(\partial^2B)^2 -\left(\phi\eth+\frac{4\beta_7}{\zeta^2}\partial^2\psi\right)\partial^2\phi\nb\\
&&-2(\psi-2\phi+2A+\alpha_1\psi\partial^2)\partial^2\psi\bigg\},
\eqn
where
\bqn\label{def}
\alpha_{1} &\equiv & \frac{8\gamma_{2} + 3\gamma_{3}}{a^2\zeta^{2}},\nb\\
\eth  &\equiv & \beta_0 + \frac{\beta_{2}+\beta_{4}}{a^2\zeta^{2}}\partial^{2}    - \frac{\beta_{8}}{a^4\zeta^{4}}\partial^4,\nb\\
\wp &\equiv &1  -  \frac{\beta_{7}}{a^2\zeta^{2}} \partial^{2}.
\eqn
Here $a$ is the scale factor of the FRW universe, which is one for the Minkowski background.
Variations of $S^{(2)}$ with respect to $A$, $\psi$, $B$, and $\phi$ yield, respectively,
\bqn
\lb{Aa}
&&\partial^2\psi=0,\\
\lb{psi}
&& \ddot{\psi} + \frac{1}{3}\partial^{2}\dot{B} + \frac{2}{3(1-3\lambda)}\left(\partial^{2}A+\partial^{2}\psi + \alpha_{1}\partial^{4}\psi\right) \nb\\
&& ~~~~~~~~ =  \frac{2}{3(1-3\lambda)}\partial^{2}\wp\phi,\\
\lb{B}
&&\big(\lambda - 1\big)\partial^{2}B = \big(1-3\lambda\big)\dot{\psi},\\
\lb{phia}
&&\eth\phi=2\wp\psi.
\eqn
 Equation.(\ref{Aa}) clearly shows that $\psi$ is not propagating, and, with proper boundary conditions, we can always set $\psi=0$.
 Similarly, Eqs.(\ref{B}), (\ref{phia})and (\ref{psi}) show that $B$, $A$, and $\phi$ are also not propagating and can be set to zero by proper boundary conditions.
 Therefore, we finally obtain
\bqn
\psi=B=A=\phi=0.
\eqn
Thus,  the scalar perturbations indeed vanish identically in the Minkowski background and, as a result,  the spin-0 gravitons are eliminated.
Then, all the problems related to the spin-0 gravitons 
disappear, including the ghost, instability, and strong coupling problems  \cite{reviews,Mukc}.

\section{Stability and Strong Coupling  of  Scalar Field}
\renewcommand{\theequation}{5.\arabic{equation}} \setcounter{equation}{0}

Since the spin-0 gravitons are eliminated, problems related to them, such as the ghost, instability, and strong coupling,
 in the gravitational sector do not exist. But, the self-interaction of matter fields
and the interaction between a matter  and a gravitational
field can still lead to strong coupling, as shown in \cite{LWWZ} for the theory with the
projectability condition. In the following,  we shall show that this is also the case here. However, it can be solved by the BPS mechanism \cite{BPSc}, by simply
introducing a new energy scale $M_{*}$, that suppresses the sixth-order spatial derivative terms. Let us first
consider the stability of a scalar field in the Minkowski background.

\subsection{Stability of  scalar field}
For a scalar field $\chi$ with the detailed balance conditions softly breaking, it is described by  \cite{BLW,WWM}
\bqn
\lb{calAa}
{\cal{L}}_M &=& {\cal{L}}_\chi^{(A,\varphi)}+{\cal{L}}_\chi^{(0)},\nb\\
{\cal{L}}_\chi^{(A,\varphi)} &=& \frac{A-{\cal{A}}}{N}\left[c_1(\chi) \Delta \chi + c_2(\chi) (\nabla \chi)^2\right] \nb\\
&&+ \frac{f}{2} [(\nabla^k \varphi)(\nabla_k \chi)]^2 \nb\\
&& - \frac{f}{N} (\dot{\chi} - N^i \nabla_i \chi) (\nabla^k \varphi) (\nabla_k \chi) ,
\\
{\cal{L}}_\chi^{(0)} &=& \frac{f}{2N^2} (\dot{\chi}-N^i \nabla_i \chi)^2 - {\cal{V}},
\eqn
where
\bqn
{\cal{V}} &=& V(\chi) + \left(\frac{1}{2}+V_1(\chi)\right)(\nabla \chi)^2 + V_2(\chi) {\cal{P}}^2_1  \nb\\
&&+ V_3(\chi) {\cal{P}}^3_1 + V_4(\chi) {\cal{P}}_2 + V_5(\chi) (\nabla \chi)^2 {\cal{P}}_2 \nb\\
&&+ V_6 {\cal{P}}_1 {\cal{P}}_2,\\
\lb{calA}
{\cal{A}}&=& - \dot{\varphi} + N^i \nabla_i \varphi + \frac{1}{2} N (\nabla_i \varphi) (\nabla^i \varphi),
\eqn
and
\bqn
{\cal{P}}_n \equiv \Delta^n \chi, \;\;\;\;\;\;\; V_6 \equiv - \sigma_3^2,
\eqn
where $\sigma_{3}$ is a constant. The coefficient $f$ in (\ref{calAa}) is a function of $\lambda$ only.
Then, it can be shown that  the Minkowski spacetime $(\bar N, \bar N^{i}, \bar g_{ij}) = (1, 0, \delta_{ij})$ is a solution of the above theory, provided that
\bq
\lb{Mink}
 \bar{A} = \bar{\varphi} = 0, \bar{\chi}= \bar{\chi}_{0}, \;\;\; V( \bar{\chi}_{0})=0=V'( \bar{\chi}_{0}),
\eq
where $ \bar{\chi}_{0}$ is a constant. Without loss of generality, we set it to zero.
Considering the perturbations (\ref{Minkp}), together with the one of the scalar field $\chi=\delta \chi$,  we find that to second order the total action  is given by
\bqn
S^{(2)}&=&\zeta^2\int dtd^3x\bigg\{(1-3\lambda)(3\dot{\psi}^2+2\dot{\psi}\partial^2B)\nb\\
& &+(1-\lambda)(\partial^2B)^2 -\left(\phi\eth+\frac{4\beta_7}{\zeta^2}\partial^2\psi\right)\partial^2\phi\nb\\
&&-2(\psi-2\phi+2A+\alpha_1\psi\partial^2)\partial^2\psi\nb\\
&&+\frac{1}{\zeta^2}\bigg[\frac{f}{2}\dot{\chi}^2-\frac{1}{2}V''\chi^2+c_1A\partial^2\chi\nb\\
&&-(\frac{1}{2}+V_1)(\partial\chi)^2-V_2(\partial^2\chi^2)^2\nb\\
&&-V_4'\chi\partial^4\chi+\sigma_3^2\partial^2\chi\partial^4\chi\bigg]\bigg\}.
\eqn
Variations of this action with respect to $A$, $\psi$, $B$, $\phi$, and $\chi$ yield,  respectively, 
\bqn
\lb{f1}
&&\psi=\frac{c_1}{4\zeta^2}\chi,\\
\lb{f2}
&&
\ddot{\psi} + \frac{1}{3}\partial^{2}\dot{B} + \frac{2}{3(1-3\lambda)}\left(\partial^{2}A+\partial^{2}\psi + \alpha_{1}\partial^{4}\psi\right) \nb\\
&& ~~~~~~~~ =  \frac{2}{3(1-3\lambda)}\partial^{2}\wp\phi,\\
\lb{f3}
&&\big(\lambda - 1\big)\partial^{2}B = \big(1-3\lambda\big)\dot{\psi},\\
\lb{f4}
&&\eth\phi=2\wp\psi,\\
\lb{f5}
&&f\ddot{\chi}+V''\chi-(1+2V_1)\partial^2\chi+2(V_2+V_4')\partial^4\chi\nb\\
&&~~~~~~~~~~~~=2\sigma_3^2\partial^6\chi+c_1\partial^2A.
\eqn
From the above  field equations, one can get a  master equation for the scalar field $\chi$, which in momentum space can be written in the form
\bqn\label{master}
\ddot{\chi}_k+\omega_k^2\chi_k=0,
\eqn
where
\bqn
\lb{omega}
\omega_k^2&=&\frac{1}{f+\frac{c_1^2}{4\zeta^2|c_\psi|^2}}\Bigg\{V''+k^2\left(1+2V_1-\frac{c_1^2}{4\zeta^2}\right)\nb\\
&&+k^4\left(2V_2+2V_4'+\frac{c_1^2}{4\zeta^2}\frac{\lambda_3}{M_A^2}\right)+2\sigma_3^2k^6\nb\\
&&+\frac{c_1^2}{4\zeta^2}\frac{2\left(1+\frac{\lambda_2}{M_A^2}k^2\right)^2k^2}{\alpha-k^2\frac{\lambda_1}{M_A^2}-k^4\frac{\lambda_4}{M_B^4}}\Bigg\},
\eqn
with
\bqn
\lambda_1 &\equiv& \frac{\beta_2+\beta_4}{\zeta^2}M_A^2,\;\;\lambda_2 \equiv \frac{\beta_7}{\zeta^2}M_A^2,\nb\\
\lambda_3 &\equiv& \frac{8\gamma_2+3\gamma_3}{\zeta^2}M_A^2,\;\;\lambda_4 \equiv \frac{\beta_8}{\zeta^4}M_B^4,\nb\\
M^{2}_{A} &\equiv& \beta^{2}\Big(2\pi G c_{1}^{2}  \frac{8\gamma_2 + 3\gamma_3}{\zeta^{2}}\ + V_{2} + V_{4}'\Big)^{-1},\nb\\
M^{4}_{B} &\equiv& \frac{\beta^{2}}{\sigma_{3}^{2}},\;\;\;
\beta^{2} \equiv \frac{2\pi G c_{1}^{2}}{|c_{\psi}|^{2}}   + \frac{f}{2},\nb\\
c_{\psi}^{2} &\equiv&  \frac{1-\lambda}{3\lambda - 1}.
\eqn
In the IR, we have $k\ll M_{\ast}={\mbox{Min}}.(M_A,\;M_B)$. Then,  we find that
\bqn
\omega_k^2=\frac{2\beta^{2}m_{\chi}^{2}}{f+\frac{c_1^2}{4\zeta^2|c_\psi|^2}}>0,
\eqn
where
\bq
\lb{mass}
m^{2}_{\chi} \equiv  \frac{1}{2\beta^{2}}V'',
\eq
denotes the mass of the scalar field. Thus,  it  is stable for $f > 0$. In the UV, we have $k^2\geq M_A, M_B$, and then we find that 
\bqn
\omega_k^2\simeq \frac{2\sigma_3^2k^6}{f+\frac{c_1^2}{4\zeta^2|c_\psi|^2}}>0,
\eqn
for $f > 0$. Therefore, in this regime the scalar field is also stabilized.  In fact, it can be made
stable in all the energy scales by properly choosing the coupling coefficients $V_{n}$, as can be seen from Eq.(\ref{omega}).

\subsection{Strong coupling of scalar field}
To study the strong coupling problem, using Eqs.(\ref{f1})-(\ref{f5}), we can integrate out $\psi,\; B,\;\phi$, and ${A}$, so
$S^{(2)}$   finally takes the form
\bqn
\lb{3.9}
S^{(2)} &=& \beta^{2}\int{dtd^{3}x\Bigg[\dot\chi^{2} - \alpha_0\big(\partial\chi\big)^{2} - m_{\chi}^{2} \chi^{2} }
 - \frac{\chi}{M_{A}^{2}} \partial^{4}\chi  \nb\\
&& + \frac{\chi}{M_{B}^{4}} \partial^{6}\chi  +\gamma\chi\partial^2\left(\frac{\wp^2\chi}{\eth}\right)\Bigg], 
\eqn
where
\bqn
\lb{3.10}
\alpha_0 &\equiv& \frac{1}{2\beta^{2}}\Big(1 + 2V_{1} - 4\pi G c_{1}^{2}\Big),\nb\\
 \gamma  &\equiv& \frac{4\pi Gc_1^2}{\beta^2}.
\eqn
As a consistency check, one can show that the variation of the action (\ref{3.9})  with respect to $\chi$ yields the master equation (\ref{master}).
In addition, when $\lambda$ satisfies the condition (\ref{1.11}), the above expression shows clearly that the scalar field is ghost free for $f > 0$ and  stable in all energy scales.

To study the strong coupling problem,   let us first note that the corresponding cubic action is given by,
  \bqn
 \lb{3.10a}
S^{(3)}&=& \int{dtd^{3}x\Bigg\{g_{1}\left(\frac{1}{\partial^{2}}\ddot{\chi}\right)\chi\partial^{2}\chi  + g_{2}\left(\frac{1}{\partial^{2}}\ddot{\chi}\right)\chi_{,i}\chi^{,i}}
 \nb\\
 && +  g_{3} \dot{\chi}^{2}\left(\frac{2\wp }{\eth} -1\right)\chi + g'_3 {\chi}\dot\chi^{2} \nb\\
 &&+ g_4 \left(\frac{\partial^{i}\partial^{j}}{\partial^{2}}\dot\chi\right)\left(\frac{\partial_{i}}{\partial^{2}}\dot{\chi}\right)\partial_j\left(\frac{2\wp }{\eth} +3\right)\chi
 \nb\\
 &&  g_5 \dot{\chi}\partial^i\left(\frac{2\wp }{\eth} -1\right)\chi\partial_{i}\left(\frac{\dot{\chi}}{\partial^{2}}\right) +  g'_{5} \dot\chi \chi^{,i}\left(\frac{\partial_{i}}{\partial^{2}}\dot\chi\right) \nb\\
&&+ g_{6}\chi^{3} + g_{7} \chi^{2}\partial^{2}\chi
 +  g_{8} \chi^{2} \partial^{4}\chi  + g_{9}  \chi^{2} \partial^{6}\chi \nb\\
 & &  + ...\Bigg\}, 
 \eqn
 where ``..." represents the fourth- and sixth-order derivative terms, which are irrelevant to the strong coupling
 problem. It also contains terms like
  \bqn
 \phi \dot{\chi}^2, \;\;
 \phi \chi^2,\;\;
 \phi \chi \partial^2 \chi,\;\;
 \phi \chi \partial^4 \chi,\;\;
 \phi \chi \partial^6 \chi,\;\;
 \chi^2 \partial^2 \phi,\;\;
 \nb\\
 \chi^2 \partial^4 \phi, \;\;
 \chi^2 \partial^6 \phi, \;\;
 \phi^2 \partial^2\chi,\;\;
 \phi^2 \partial^4 \chi,\;\;
 \phi^2 \partial^6 \chi, \;\;
 \phi \chi \partial^2 \phi
 \nb\\
 \phi \chi \partial^4 \phi,\;\;
 \phi \chi \partial^6 \phi,\;\;
 \phi^2 \partial^2 \phi,\;\;
 \phi^2 \partial^4 \phi,\;\;
 \phi^2 \partial^6 \phi, \;\; \dots.
 \nb\\
 \eqn
Since these terms are also independent of $\lambda$, they are irrelevant to the strong coupling
 problem, too. The coefficients $g_{s}$ are defined as
 \bqn
 \lb{3.10ba}
 g_{1} &=& \frac{c_1^3}{8\zeta^4|c_\psi|^{2}},\;\;\;
 g_{2} =  \frac{1}{|c_\psi|^{2}}\left(\frac{5c_1^3}{32\zeta^4}-\frac{c_1c_2}{4\zeta^2}\right),\nb\\
 g_{3} &=& - \frac{c_1^3}{32\zeta^4 |c_\psi|^{2}},\;\; g'_3= -\frac{3fc_1}{8\zeta^2},\;\;\; g_{4} =  \frac{c_1^3}{64\zeta^4 |c_\psi|^{4}}, \nb\\
 g_{5} &=& -\frac{c_1^3}{64\zeta^4|c_\psi|^{4}} ,\;\; g'_5= \frac{c_1f}{4\zeta^2 |c_\psi|^{2}}, \;\;\;
 g_{6} = \frac{3c_1}{8\zeta^2}\ddot{V}-\frac{\dddot{V}}{6}, \nb\\
 g_{7} &=&
 \frac{\dot{V}_1}{2}+\frac{c_1^2c_2}{8\zeta^2}-\frac{c_1}{16\zeta^2}-\frac{c_1V_1}{8\zeta^2}, \nb\\
 g_{8} &=& \tilde{A}_{1}(\gamma_2, \gamma_3, c_1)     + {B}_{1}\big(V_2, V_4\big), \nb\\
 g_{9} &=& \tilde{A}_{2}(\gamma_5, c_1)  + {B}_{2}\big(V_3, V_5, V_6\big),
 \eqn
 where  $\tilde{A}_{i} \equiv (4\pi G c_{1})^{3}A_{i}$. 
 Depending on the energy scales, each of these terms will have different scalings. Thus,
  in the following  we consider them separately.

 \subsubsection{$|\nabla| \ll M_{*}$}

When $|\nabla| \ll M_{*}$, where $M_{*} = {\mbox{Min.}}\big(M_{A}, M_{B}\big)$,
  we find that
\bq
\lb{3.19a}
\eth \simeq \beta_{0},\;\;\; \wp \simeq 1, \;\;\;
\phi \simeq  \frac{2}{ \beta_{0}} \psi.
\eq
Then,  Eq.(\ref{3.9}) reduces to
 \bq
\lb{3.11} S^{(2)} \simeq \beta^{2}\int{dtd^{3}x\Big[\dot\chi^{2} -
\tilde{\alpha}\big(\partial\chi\big)^{2}\Big]},
\eq
where
\bqn
\tilde{\alpha}=\alpha_0+\frac{\gamma}{ \beta_{0}}.
\eqn
Note that in writing the above expression, without loss of generality, we had assumed that $|\nabla| \gg m_{\chi}$.
By setting
\bq
\lb{3.12}
t = b_{1}\hat{t},\;\;\; x^{i} = b_{2}\hat{x}^{i},
\;\;\; \chi= b_{3}\hat\chi,
\eq
  Eq.(\ref{3.11}) can be brought into the ``canonical" form,
\bq
\lb{3.13}
S^{(2)} \simeq
\int{d\hat{t}d^{3}\hat{x}\Big[\big(\hat\chi^{*}\big)^{2} -
\big(\hat{\partial}\hat{\chi}\big)^{2}\Big]},
\eq
in which the coefficient of each term is order of $1$, for
\bq
\lb{3.14}
b_{2} = b_{1}\sqrt{\tilde{\alpha}},\;\;\; b_{3} =\frac{1}{b_{1}\beta\tilde{\alpha}^{3/4}},
\eq
where $\hat\chi^{*} \equiv d\hat\chi/d\hat{t}$. Note that the requirement that the coefficient of each term be  order of $1$ is
important in order to obtain a correct coupling strength \cite{BPS,HW,LWWZ}.

When $|\nabla| \ll M_{*}$, the third-order action (\ref{3.10a}) can be expressed as
 \bqn
 \lb{3action2}
S^{(3)}&=& \int{dtd^{3}x\Bigg\{g_{1}\left(\frac{1}{\partial^{2}}\ddot{\chi}\right)\chi\partial^{2}\chi  + g_{2}\left(\frac{1}{\partial^{2}}\ddot{\chi}\right)\chi_{,i}\chi^{,i}}
 \nb\\
 &&  +\hat g_3 {\chi}\dot\chi^{2}
 + \hat g_4 \left(\frac{\partial^{i}\partial^{j}}{\partial^{2}}\dot\chi\right)\left(\frac{\partial_{i}}{\partial^{2}}\dot{\chi}\right)\partial_j\chi
 \nb\\
 && + \hat g_{5} \dot\chi \chi^{,i}\left(\frac{\partial_{i}}{\partial^{2}}\dot\chi\right) + g_{6}\chi^{3} + g_{7} \chi^{2}\partial^{2}\chi
 +  g_{8} \chi^{2} \partial^{4}\chi \nb\\
 && + g_{9}  \chi^{2} \partial^{6}\chi  + ...\Bigg\},
 \eqn
where $g_{s}$ are  given by Eq.(\ref{3.10ba}), and 
\bqn
\hat g_3&=&\left(1-\frac{2}{\beta_0}\right)\frac{c_1^3}{32\zeta^4 |c_\psi|^{2}}-\frac{3fc_1}{8\zeta^2},\nb\\
\hat  g_4&=&\left(3+\frac{2}{\beta_0}\right)\frac{c_1^3}{64\zeta^4 |c_\psi|^{4}},\nb\\
\hat g_5&=&\left(1-\frac{2}{\beta_0}\right)\frac{c_1^3}{64\zeta^4|c_\psi|^{4}}+\frac{c_1f}{4\zeta^2 |c_\psi|^{2}}.
\eqn
Inserting Eq.(\ref{3.12})   into Eq.(\ref{3action2}), we obtain
  \bq
 \lb{3.10bb}
 S^{(3)} = \frac{1}{b_{1}\beta^{3}\beta_0^{3/4}} \hat{S}^{(3)},
 \eq
where
\bqn
\lb{3.10b}
  \hat{S}^{(3)} &\equiv&
 \int{d\hat{t}d^{3}\hat{x}\Bigg\{g_{1}\left(\frac{1}{\hat\partial^{2}}\hat{\chi}^{**}\right)\hat\chi\hat\partial^{2}\hat\chi}\nb\\
 && + g_{2}\left(\frac{1}{\hat\partial^{2}}\hat{\chi}^{**}\right)\hat\partial_{i}\hat\chi\hat\partial^{i}\hat\chi
 +  \hat g_{3} {\hat\chi}\hat\chi^{*2}
 \nb\\
 &&  + \hat g_{4}\left(\hat\partial_{i}\hat\chi\right)\left(\frac{\hat\partial_{j}}{\hat\partial^{2}}\hat\chi^{*}\right)\left(\frac{\hat\partial^{i}\hat\partial^{j}}{\hat\partial^{2}}\hat\chi^{*}\right)
 \nb\\
 &&  +  \hat g_{5} \hat\chi^{*} \left(\frac{\hat\partial_{i}}{\hat\partial^{2}}\hat\chi^{*}\right)\hat\partial_{i}\hat\chi\nb\\
 &&  +g_{6} b_{1}^{2}\hat\chi^{3} + \frac{g_{7}b_{1}^{2}}{b_{2}^{2}} \hat\chi^{2}\hat\partial^{2}\hat\chi  \nb\\
 && +   \frac{g_{8}b_{1}^{2}}{b_{2}^{4}} \hat\chi^{2} \hat\partial^{4}\hat\chi  +  \frac{g_{9}b_{1}^{2}}{b_{2}^{6}} \hat\chi^{2} \hat\partial^{6}\hat\chi \nb\\
 & &  + ...\Bigg\}.
 \eqn

On the other hand, from Eq.(\ref{3.13}) one finds  that $S^{(2)}$ is invariant under  the rescaling,
\bq
\lb{3.15}
\hat{t} \rightarrow b^{-1}\hat{t}, \;\;\; \hat{x}^{i} \rightarrow b^{-1}\hat{x}^{i},\;\;\; \hat{\chi} \rightarrow b\hat{\chi}.
\eq
Then, it can be shown that  the  terms of $g_{1, 2, ..., 5}$ and $g_{7}$  in $S^{(3)}$ all scale as $b$, while the terms of $g_{6, 8, 9}$ scale as $b^{-1},\; b^{3}, \; b^{5}$, respectively.
Therefore, except for the $g_{6}$ term, all the others are irrelevant and nonrenormalizable \cite{Pol}. For example, considering a process with  an energy $E$, then we
find that the fourth term has the contribution
\bq
\lb{3.16}
\int{d\hat{t}d^{3} \hat{x}\left(\hat\partial_{i}\hat\chi\right)\left(\frac{\hat\partial_{j}}{\hat\partial^{2}}\hat\chi^{*}\right)\left(\frac{\hat\partial^{i}\hat\partial^{j}}{\hat\partial^{2}}\hat\chi^{*}\right)}
\simeq E.
\eq
Since the action $S^{(3)}$ is dimensionless, we must have
\bqn
\lb{3.17}
 && \frac{\lambda_{4}}{b_{1}\beta^{3}\beta_0^{3/4}} \int{d\hat{t}d^{3} \hat{x}\left(\hat\partial_{i}\hat\chi\right)\left(\frac{\hat\partial_{j}}{\hat\partial^{2}}\hat\chi^{*}\right)
 \left(\frac{\hat\partial^{i}\hat\partial^{j}}{\hat\partial^{2}}\hat\chi^{*}\right)}
\nb\\
&& ~~~~~~~~~~~~~~ \simeq \frac{E}{\Lambda^{(4)}_{SC}},
\eqn
where $\Lambda^{(4)}_{SC}$ has the same dimension of  $E$, and is given by
\bq
\lb{3.18}
\Lambda^{(4)}_{SC} = \frac{b_{1}\beta^{3}\tilde{\beta_0}^{3/4}}{g_{4}}.
\eq
Similarly, one can find $\Lambda^{(n)}_{SC}$ for all the other nonrenormalizable terms. But,  when $\lambda \rightarrow 1$ (or $c_{\psi} \rightarrow
0$),  the lowest one of the $\Lambda^{(n)}_{SC}$'s  is given by $\Lambda^{(4)}_{SC}$, so we have
\bq
\lb{3.19b}
\Lambda_{\hat{\omega}}  \equiv \frac{b_{1}\beta^{3}\tilde{\beta_0}^{3/4}}{g_{4}},
\eq
above which the nonrenormalizable $\hat g_{4}$ term becomes larger than unity, and the process runs into the strong coupling regime.
Back to the physical coordinates  $t$ and $x$, the corresponding energy and momentum scales are given, respectively, by
\bqn
\lb{3.20}
\Lambda_{\omega} &=& \frac{\Lambda_{\hat{\omega}}}{b_{1}} 
\simeq {\cal{O}}(1)\left(\frac{\zeta}{c_1}\right)^{3/2}M_{pl}\left|c_{\psi}\right|^{5/2},\nb\\
\Lambda_{k} &=& \frac{\Lambda_{\hat{\omega}}}{b_{2}} 
\simeq {\cal{O}}(1)\left(\frac{\zeta}{c_1}\right)^{1/2}M_{pl}\left|c_{\psi}\right|^{3/2}.
\eqn
In particular, for $c_1 \simeq \zeta$,
we find that $\Lambda_{\omega}  \simeq M_{pl}\left|c_{\psi}\right|^{5/2}$, which is precisely the result obtained in \cite{HW}.

\begin{figure}[tbp]
\centering
\includegraphics[width=8cm]{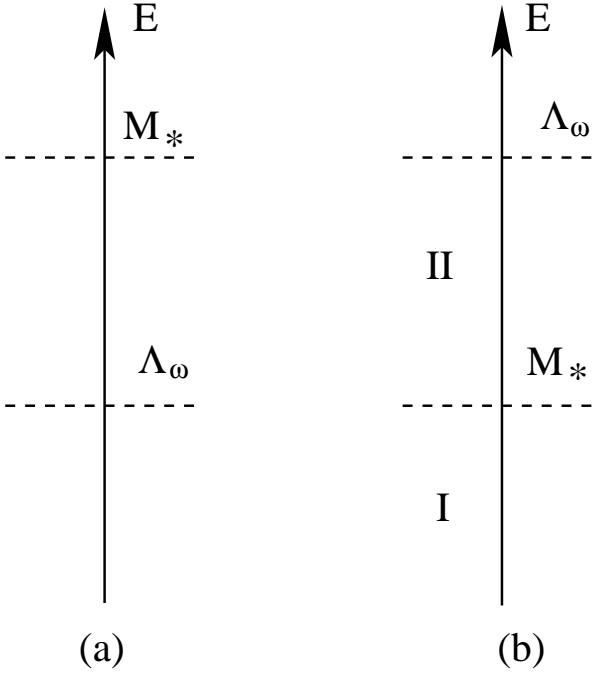}
\caption{The energy scales: (a) $\; \Lambda_{\omega} <  M_{*}$; and (b) $\; \Lambda_{\omega} > M_{*}$.}
\label{fig1}
\end{figure}

It should be noted that the above conclusion is true only for $M_{*} > \Lambda_{\omega}$, that is,
\bq
\lb{3.21}
M_{*} > \left(\frac{\zeta}{c_1}\right)^{3/2}M_{pl}\left|c_{\psi}\right|^{5/2},
\eq
as shown by Fig. \ref{fig1}(a).

When $M_{*} < \Lambda_{\omega}$, the above analysis holds only  for the processes with $E \ll M_{*}$ [Region I in Fig.\ref{fig1}(b)]. However,
 when   $E \gtrsim M_{*}$ and before the strong coupling energy scale $\Lambda_{\omega}$
 reaches [cf. Fig.\ref{fig1}(b)], the high-order derivative terms of $M_{A}$ and $M_{B}$  in Eq.(\ref{3.9}) cannot be
 neglected any more, and one has to take these terms into account. It is exactly because the presence
 of these terms that the strong coupling problem is cured \cite{BPSc}. In the following,  we show that this is also the case here.

 \subsubsection{$M_{*} < \Lambda_{\omega},\;\;\; M_{A} < M_{B}$}

When $M_{A} < M_{B}$, we have $M_{*} = M_{A}$. In this case, we find that
\bq
\lb{3.19}
\eth \simeq \frac{\lambda_1}{M_A^2}\partial^2,\;\;\; \wp \simeq -\frac{\lambda_3}{M_A^2}\partial^2, \;\;\;
\phi \simeq  -\frac{2\lambda_3}{\lambda_1} \psi.
\eq
For the processes with $E \gtrsim M_{A}$,  Eq.(\ref{3.9}) reduces to
\bqn
\lb{3.22}
S^{(2)} &=& \beta^{2}\int{dtd^{3}x\left(\dot\chi^{2}    -  \frac{1}{\mu_{A}^{2}} \chi\partial^{4}\chi\right)},\\
\frac{1}{\mu_A^2}&=&\left(1-\frac{\gamma\lambda_2}{\lambda_3}\right)\frac{1}{M_A^2},
\eqn
and the coefficients $\hat g_3$, $ \hat g_4$, and $ \hat g_5$ now are defined as
\bqn
\hat g_3&=&\left(1+\frac{2\lambda_3}{\lambda_1}\right)\frac{c_1^3}{32\zeta^4 |c_\psi|^{2}}-\frac{3fc_1}{8\zeta^2},\nb\\
\hat g_4&=&\left(3-\frac{2\lambda_3}{\lambda_1}\right)\frac{c_1^3}{64\zeta^4 |c_\psi|^{4}},\nb\\
\hat g_5&=&\left(1+\frac{2\lambda_3}{\lambda_1}\right)\frac{c_1^3}{64\zeta^4|c_\psi|^{4}}+\frac{c_1f}{4\zeta^2 |c_\psi|^{2}}.
\eqn
Note that to have $\mu_{A}$ real, we must assume that 
\bqn
1-\frac{\gamma\lambda_2}{\lambda_3}>0.
\eqn
To study the strong coupling problem, we shall follow what we did in the last case, by first  writing $S^{(2)}$ in its canonical form,
\bq
\lb{3.22a}
S^{(2)} =  \int{d\hat{t}d^{3}\hat{x}\left(\hat\chi^{*2}    - \hat\chi\hat\partial^{4}\hat\chi\right)},
\eq
 through the transformations (\ref{3.12}).  It can be shown that now $b_{2}$ and $b_{3}$ are given by
\bq
\lb{3.23}
b_{2} = \sqrt{\frac{b_{1}}{\mu_{A}}},\;\;\;
b_{3} = \frac{\mu_{A}^{3/4}}{b_{1}^{1/4}\beta},
\eq
for which the cubic action $S^{(3)}$ takes the form
  \bq
 \lb{3.10c}
 S^{(3)} = \frac{\mu_{A}^{3/4}}{b_{1}^{1/4}\beta^{3}} \hat{S}^{(3)},
   \eq
where $\hat{S}^{(3)}$ is given by Eq.(\ref{3.10b}).
Because of the nonrelativistic nature of the action (\ref{3.22a}), its scaling becomes anisotropic,
\bq
\lb{3.24}
\hat{t} \rightarrow b^{-2}\hat{t},\;\;\;
\hat{x}^{i} \rightarrow b^{-1}\hat{x}^{i},\;\;\;
\hat{\chi} \rightarrow b^{1/2}\hat{\chi}.
\eq
Then, we find that   the first five terms in Eqs.(\ref{3.10c}) and  (\ref{3.10b}) scale as $b^{1/2}$, while the terms of $g_{6, ...,9}$ scale, respectively, as $b^{-7/2},\;
b^{-3/2},\; b^{1/2},\; b^{5/2}$. Thus, except for the $g_{6}$ and $g_{7}$ terms, all the others are not renormalizable. It can be also shown that the processes
with energy higher than $\Lambda_{\omega}^{(A)}$ become strong coupling, where $\Lambda_{\omega}^{(A)}$ is given by
\bq
\lb{3.25}
\Lambda_{{\omega}}^{(A)}  \simeq \left(\frac{M_{pl}}{\mu_{A}}\right)^{3}M_{pl}\left|c_{\psi}\right|^{4},\; (M_{A} < M_{B}).
\eq
Therefore, when the fourth-order derivative terms dominate, the strong coupling problem still exists. This is expected, as power counting tells us that the theory is
renormalizable only when $z \ge 3$ [cf. Eq.(\ref{1.1})]. Indeed, as will be shown
below, when the sixth-order spatial derivative terms dominate, the strong coupling problem does not exist.

\subsubsection{$M_{*} < \Lambda_{\omega},\;\;\; M_{A} \gtrsim M_{B}$}

In this case,  we have $M_{*} = M_{B}$, and   for  processes with $E \gtrsim M_{B}$,   Eq.(\ref{3.9}) reduces to
\bq
\lb{3.26}
S^{(2)} = \beta^{2}\int{dtd^{3}x\left(\dot\chi^{2}    -  \frac{1}{M_{B}^{4}} \chi\partial^{6}\chi\right)}.
\eq
Then, all the terms which contain $\phi$ in (\ref{3.10a}) can be neglected, and  the coefficients $\hat g_3$, $\hat g_4$, and $\hat g_5$ in Eq.(\ref{3action2}) now become,
\bqn
\hat g_3&=&\frac{c_1^3}{32\zeta^4 |c_\psi|^{2}}-\frac{3fc_1}{8\zeta^2},\nb\\
\hat g_4&=&\frac{c_1^3}{64\zeta^4 |c_\psi|^{4}},\nb\\
\hat g_5&=&\frac{c_1^3}{64\zeta^4|c_\psi|^{4}}+\frac{c_1f}{4\zeta^2 |c_\psi|^{2}}.
\eqn
Then, by the transformations (\ref{3.12}) with
\bq
\lb{3.27}
b_{2} = \frac{b_{1}^{1/3}}{M_{B}^{2/3}},\;\;\;
b_{3} = \frac{M_{B}}{\beta},
\eq
 we obtain,
\bq
\lb{3.28}
S^{(2)} =  \int{d\hat{t}d^{3}\hat{x}\left(\hat\chi^{*2}    - \hat\chi\hat\partial^{6}\hat\chi\right)},
\eq
while the cubic action $S^{(3)}$ becomes,
  \bq
 \lb{3.10d}
 S^{(3)} = \frac{M_{B}}{\beta^{3}} \hat{S}^{(3)}.
   \eq
Equation (\ref{3.28}) is invariant under the rescaling,
\bq
\lb{3.29}
\hat{t} \rightarrow b^{-3}\hat{t},\;\;\;
\hat{x}^{i} \rightarrow b^{-1}\hat{x}^{i},\;\;\;
\hat{\chi} \rightarrow \hat{\chi}.
\eq
Then, it can be shown that   the first five terms in Eqs.(\ref{3.10d}) and (\ref{3.10b}) are scaling-invariant, and so the last term. The terms of $g_{6, 7, 8}$, on the other hand,
 scale, respectively, as $b^{-6},\;
b^{-4},\; b^{-2}$. Therefore, the first five  terms as well as  the last one now all become strictly renormalizable, while the   $g_{6},\; g_{7}$ and $g_{8}$ terms become superrenormalizable
\cite{Pol}. To have these strictly renormalizable terms be weakly coupling, we require their coefficients be less than unity,
\bq
\lb{3.30}
\frac{M_{*}}{\beta^{3}} g_{n} < 1, \; (n = 1, ..., 5, 9).
\eq
For $g \sim 1$ (or $|c_{\psi}| \sim 0$), we find that the above condition holds for
$$
M_{*} < \frac{2}{3} M_{pl}\left|c_{\psi}\right|.
$$
It can be shown that this condition holds identically, provided that $M_{*} < \Lambda_{\omega}$, that is,
\bq
\lb{3.32}
 M_{*} <  \left(\frac{\zeta}{c_1}\right)^{3/2}M_{pl}\left|c_{\psi}\right|^{5/2}.
\eq
[Recall $\Lambda_{\omega}$ is given by  Eq.(\ref{3.20}) and $M_{*} = M_{B}$.]
One can take $c_{1} \simeq M_{pl}$, but now a more reasonable choice is   $c_1 \simeq M_{*}$. Then, the condition (\ref{3.32}) becomes
 \bq
\lb{3.31}
M_{*} <   M_{pl}\left|c_{\psi}\right|^{1/2},\; (c_1 = M_{*}),
\eq
 which is much less restricted than the one of   $c_1 \simeq M_{pl}$. 
In addition, in order to have the sixth-order derivative terms dominate, we must also require 
\bq
\lb{3.33}
 M_{A} \gtrsim M_{*}.
\eq
Therefore, it is concluded that, {\em provided that  conditions (\ref{3.32}) and (\ref{3.33}) hold,   the extended version of the HL gravity with the detailed balance condition softly breaking
 but without the projectability condition  is absent of the strong coupling problem}.

\section{Cosmological Models and the Flatness Problem}
\renewcommand{\theequation}{6.\arabic{equation}} \setcounter{equation}{0}

One of the main motivations of inflation was to solve the horizon and flatness problems, encountered in the standard Big Bang model \cite{Inflation}. In the HL theory, the anisotropic
scaling (\ref{1.1})  provides a solution to the horizon problem and generation of scale-invariant perturbations even without inflation \cite{Mukohyama:2009gg}. Clearly, these statements
are also true in our current setup developed above.  In this section, we    shall  show that   the homogeneous and isotropic universe is also necessarily
flat, when the enlarged 
symmetry (\ref{symmetry})  is introduced. This was first noted for a scalar field  \cite{HW}. Here we argue that it is
true for all the viable cosmological models. To this purpose, let us consider the  general FRW universe,
\bqn
\lb{metricB}
ds^2&=&a^2(-d\eta^2+\gamma_{ij}dx^idx^j),\nb\\
\gamma_{ij}&=&\frac{\delta_{ij}}{(1+kr^2/4)^2},
\eqn
where
\bq
\delta_{ij}=\cases{
                1, &  $i=j$, \cr
                0, &$i \not= j$.}
   \eq
Then,  we have
\bqn
\hat{\Gamma}^k_{ij}&=&-\frac{1}{2}\left(\frac{k}{1+\frac{1}{4}kr^2}\right)\left(x_i\delta_j^k+x_j\delta_i^k-x^k\delta_{ij}\right),\nb\\
\hat{R}_{ij}&=&2k\gamma_{ij},\;\;\;
\hat{K}_{ij}=-a{\cal{H}}\gamma_{ij}, 
\eqn
where ${\cal{H}}=a'/a$. We use symbols with hats to denote  quantities of the background in the conformal coordinates (\ref{metricB}), following the conventions
given in \cite{HW,BLW,LWWZ}. Using the $U(1)$ gauge freedom of Eqs.(\ref{gaugeAA}) and (\ref{eqb}), we can always set one of $\hat A$ and $\hat \varphi$ to zero. In this paper, we choose the gauge
\bqn
\lb{gaugevarphi}
\hat{\varphi}(\eta)=0.
\eqn
Then, we find that
\bqn
\hat{{\cal{L}}}_K &=& (1-3\lambda) \frac{3 {\cal{H}}^2}{a^2},\nb\\
\hat{{\cal{L}}}_V &=& 2 \Lambda - \frac{6 k}{a^2} + \frac{3 \gamma_2 + \gamma_3}{\zeta^2} \frac{12 k^2} {a^4},\nb\\
\hat{\pi}^{ij} &=& (1-3\lambda) \frac{{\cal{H}}}{a}g^{ij},\nb\\
\hat{{\cal{L}}}_A &=& - \frac{\hat{A}}{a} \left(\frac{6k}{a^2}- 2 \Lambda_g\right),\nb\\
\hat{{\cal{L}}}_\varphi &=& \hat{F}^N = \hat{F}^\varphi_{ij} = \hat{F}^a_{ij} = 0, \nb\\
\hat{F}_{ij} &=& -a^2 \Big[\Lambda - \frac{k}{a^2} - \frac{2k^2}{a^4} \frac{3 \gamma_2 + \gamma_3}{\zeta^2}\Big] \gamma_{ij}.
\eqn
Because of the spatial homogeneity, both $\hat{{\cal{L}}}_K$ and $\hat{{\cal{L}}}_V$
are independent of the spatial coordinates, and the matter sector takes the forms,
\bqn
\hat{J}^t = -2 \hat{\rho}, \; \hat{J}^i = 0, \; \hat{\tau}_{ij} = \hat{p} \hat{g}_{ij},
\eqn
where $\hat{\rho}$ and $\hat{p}$ denote the total energy density and pressure, respectively. Then the Hamilton constraint (\ref{hami}) reduces to the super-Hamiltonian constraint
$\hat{{\cal{L}}}_K + \hat{{\cal{L}}}_V = 8 \pi G \hat{J}^t$ \footnote{Since now the Hamiltonian constraint is local, one cannot include a ``dark matter component as
an integration constant," as in the case with the projectability condition \cite{Mukohyama:2009gg}.}, which leads to the modified Friedmann Equation,
\bqn
\frac{3 \lambda - 1}{2}\frac{{\cal{H}}^2}{a^2} +\frac{ k}{a^2}= \frac{8 \pi G}{3} \hat{\rho} + \frac{\Lambda}{3}  + \frac{3 \gamma_2 + \gamma_3}{\zeta^2} \frac{2 k^2} {a^4}.
\eqn
It can be shown that the supermomentum constraint (\ref{mom}) is satisfied identically, while  Eq.(\ref{phi}) and Eq.(\ref{A}) give, respectively,
\bqn
\lb{eq6a}
\label{var}\frac{{\cal{H}} }{a} \left(\Lambda_g - \frac{k}{a^2}\right) = - \frac{8 \pi G}{3} \hat{J}_\varphi,\\
\lb{eq6b}
\label{bA}\frac{k}{a^2} - \frac{\Lambda_g}{3} = \frac{4\pi G \hat{J}_A}{3}.
\eqn
The dynamical equation (\ref{dyn}), on the other hand, reduces to
\bqn
\frac{1-3\lambda}{2} \left( 2 {\cal{H}}' + {\cal{H}}^2\right) + a \hat{A} \left(\frac{k}{a^2}-\Lambda_g\right) \nb\\
+a^2 \left(  \Lambda - \frac{k}{a^2} - \frac{2k^2}{a^4} \frac{3 \gamma_2 + \gamma_3}{\zeta^2}\right)= 8 \pi G \hat{p} a^2.
\eqn
The conservation law of momentum (\ref{mom conservation}) is satisfied
identically, while the one of energy (\ref{energy conservation}) reduces to,
\bqn\label{conservation-frw}
\hat{\rho}'+3 {\cal{H}} (\hat{\rho} + \hat{p}) =  \hat{A} \hat{J}_\varphi .
\eqn

It is remarkable to note that when
\bqn
\hat{J}_A=\hat{J}_\varphi = 0,
\eqn
Eqs.(\ref{eq6a}) and (\ref{eq6b}) show that the universe is necessarily flat,
\bq
\lb{eq6c}
k = 0 = \Lambda_{g}.
\eq
As first noted in \cite{HW}, this is true for the universe dominated by a single scalar field. 

In general, the coupling of the gauge field $A$ and the Newtonian prepotential $\varphi$ to a matter field $\psi_{n}$ is given by
\cite{Silva},
\bq
\lb{coupling}
\int{dt d^{3}x \sqrt{g} Z(\psi_{n},  g_{ij}, \nabla_{k})(A - {\cal{A}})},
\eq
where ${\cal{A}}$ is defined in Eq.(\ref{calA}),
 and $Z$ is the most general scalar operator under the full symmetry of Eq.(\ref{symmetry}),
with its dimension $[Z] = 2$. For a single scalar field, $Z(\chi,  g_{ij}, \nabla_{k})$ is given by
$Z(\chi,  g_{ij}, \nabla_{k}) =   c_1 \Delta \chi + c_2 (\nabla \chi)^2$, as one can see from Eq.(\ref{calAa}).
In the   multi-scalar field case,  $Z$ takes the form
\bqn
\lb{eq6d}
Z &=& \sum_{i =1}^{ n}{c^{(i)}_1 \Delta \chi^{(i)}}\nb\\
&&  +  \sum_{i,  j =1}^{ n}{c_{(i, j)}(\nabla \chi^{(i)}) (\nabla \chi^{(j)})},
\eqn
for which we have $\bar{J}_{A} = 0 = \bar{J}_{\varphi}$ with the gauge (\ref{gaugevarphi}). Thus, in the case of multi-scalar fields,
the universe is necessarily flat, too.

For a vector field ($A_{0}, A_{i}$), we have $[A_{0}] = 2,\; [A_{i}] = 0$ \cite{KKa,BLWb}. Then, we find
\bq
\lb{eq6e}
Z(A_{0}, A_{i}, g_{ij}, \nabla_{k}) = {\cal{K}}B_{i}B^{i},
\eq
where ${\cal{K}}$ is an arbitrary function of $A^{i}A_{i}$, and
\bq
\lb{eq6f}
  B_i= \frac{1}{2}\frac{\varepsilon_i^{\;\;jk}}{\sqrt{g}}{\cal{F}}_{jk},\;\;\;
  \nabla^iB_i=0,
  \eq
  with ${\cal{F}}_{ij} \equiv \partial_{j}A_{i} - \partial_{i}A_{j}$.  This can be easily generalized to several vector fields,
  $(A^{(n)}_{0}, A^{(n)}_{i}$), for which we have
  \bq
\lb{eq6g}
Z(\vec{A}_{0}, \vec{A}_{i}, g_{ij}, \nabla_{k}) = \sum_{m,n }{\cal{K}}_{mn}B^{(m)}_{i}B^{(n) i},
\eq
where ${\cal{K}}_{m n}$ is an arbitrary function of $A^{(k) i}A^{(l)}_{i}$. Then,   in the FRW background, we
have $\bar{J}_{A} = 0$, because $\bar{B}^{(m)}_{i} = 0$ \cite{GMV}. With the gauge choice (\ref{gaugevarphi}), it is easy to show that
$ \bar{J}_{\varphi} = 0$, too.
 Therefore,  an early universe dominated by  vector
fields is also necessarily  flat. This can be further generalized to the case of Yang-Mills fields \cite{CH}.

For fermions, on the other hand,
their dimensions are $[\psi_{n}] = 3/2$ \cite{Alex}. Then, $Z(\psi_{n}, g_{ij}, \nabla_{k})$ cannot be a functional of $\psi_{n}$. Therefore,
in this case $\bar{J}_{A}$ and $\bar{J}_{\varphi}$ vanish identically.

In review of the above, it is not difficult to argue that, with the special form of the coupling given by Eq.(\ref{coupling}),
  {\em the universe is necessarily  flat for all the cosmologically viable models in the current setup}.

  Similar conclusion is also obtained in the case
  with the projectability condition \cite{HWb}.  Therefore, in the rest of this paper, we shall consider only the flat FRW universe.



\section{Cosmological Perturbations}
\renewcommand{\theequation}{7.\arabic{equation}} \setcounter{equation}{0}

In this section, we consider the linear perturbations in a flat FRW universe.  Let us first write the linear perturbations in the form
\cite{WM,Wang,HW}
\bqn
\lb{pert}
\delta N&=& a \phi,\;\;\; \delta N_i= a^2 (\partial_i B-S_i),\nb\\
\delta g_{ij}&=& a^2 \left(-2 \psi \delta_{ij} + 2 \partial_i \partial_j E +2\partial_{(i} F_{j)}+H_{ij}\right),\nb\\
A&=& \hat{A}+\delta A,\;\;\;
\varphi = \hat{\varphi}+\delta \varphi,
\eqn
with $\hat{\varphi} = 0$, as one can see from Eq.(\ref{gaugevarphi}),  and
\bq
\lb{eq7ab}
\partial^{i}S_{i} =  \partial^{i}F_{i} = H^{i}_{\;\;i} = 0,\;\;\; \partial^{i}H_{ij } = 0.
\eq
In the following, we shall consider the scalar, vector and tensor perturbations separately.

\subsection{Scalar Perturbations}

For the scalar perturbations  $\phi,\;B,\;\psi,\;E$, we choose the quasilongitudinal gauge \cite{WM},
\bqn
\lb{eq7.0}
E=\delta\varphi=0.
\eqn
Then, to first order we find that
\bqn
\sqrt{g}&=&(1-3\psi)a^3,\\
\delta \Gamma_{ij}^k&=& -(\delta_i^k \partial_j \psi+\delta_j^k \partial_i \psi-\delta_{ij}\partial^k\psi),\\
\delta R_{ij}&=& \delta_{ij} \partial^2\psi+\partial_i \partial_j \psi,\\
\delta R &=& 2\psi \hat{R}+\frac{4 \partial^2\psi}{a^2},\\
\delta K_{ij}&=& a\Big[(\phi{\cal{H}}+2\psi {\cal{H}} +\psi')\delta_{ij}+\partial_i\partial_j B\Big],\\
\delta K &=& \frac{1}{a} (3\phi {\cal{H}}+\partial^2 B +3\psi').
\eqn
Other useful quantities are given in Appendix B. Thus,  the field equations Eq.(\ref{phi}), Eq.(\ref{A}), the momentum constraint (\ref{mom}),
the Hamiltonian constraint (\ref{hami}), the trace  and traceless parts of dynamical equation (\ref{dyn}) are given, respectively, by
\bqn
\lb{eq7a}
&& \partial^2 \Big[2{\cal{H}} (\psi-\phi) + (1-\lambda) (\partial^2 B + 3\psi'+3{\cal{H}}\phi)\Big]\nb\\
&& ~~~~~ =8\pi G a^3 \delta J_\varphi, \\
\lb{eq7b}
&& \partial^2 \psi = 2 \pi G a^2 \delta J_A, \\
\lb{eq7c}
&& (3\lambda -1)(\psi' +\phi {\cal{H}}) +(\lambda-1) \partial^2 B=8\pi G a q,\\
\lb{eq7d}
&& \frac{3\lambda-1}{2} {\cal{H}} \left(3 \psi' + 3 \phi {\cal{H}} +\partial^2 B\right) \nb\\
&&~~~~~   - \wp \partial^2\psi+\frac{1}{2} \eth \partial^2\phi =- 4 \pi G a^2 \delta \mu, \\
\lb{eq7e}
&& \psi''+ 2{\cal{H}} \psi' + {\cal{H}}\phi'  +\left(2{\cal{H}}'+{\cal{H}}^2\right)\phi\nb\\
&& ~~~~~  +\frac{\lambda -1}{3\lambda-1}\partial^{2}\left(\psi +  \alpha_1 \partial^2 \psi - \wp \phi
 - \frac{\hat{A}\psi- \delta A}{a}\right)\nb\\
%
&& ~~~~~  =\frac{8\pi G a^2}{3\lambda-1}\left(\delta {\cal{P}} + \frac{3\lambda -1}{3}\partial^{2}\Pi\right) ,\\
\lb{eq7f}
&& B'+2{\cal{H}}B - \psi+\wp \phi   \nb\\
&&~~~~  - \alpha_1 \partial^2 \psi +\frac{\hat{A} \psi-\delta A}{a}  =-8\pi G a^2 \Pi,
\eqn
where
\bqn
\delta \mu &\equiv& - \frac{1}{2} \delta J^t,\;\;\; \delta J^i \equiv \frac{1}{a^2} \partial^i q,\nb\\
\delta \tau_{ij}& \equiv& a^2 \big[(\delta {\cal{P}}-2 \hat{p} \psi)\delta_{ij} + \Pi_{,<ij>}\big],\nb\\
\Pi_{,<ij>} &\equiv&  \Pi_{,ij} - \frac{1}{3} \delta_{ij} \partial^2 \Pi.
\eqn
In the above equations,   $\alpha_1,\;\wp$ and $ \eth$ are defined by Eq.(\ref{def}).
The conservation laws (\ref{energy conservation}) and (\ref{mom conservation}) to first order now read,
\bqn
\lb{eq7g}
&& \int d^3 x \Big\{\delta \mu'+ 3 {\cal{H}} (\delta \mu + \delta {\cal{P}}) - 3 (\hat{\rho}+\hat{p}) \psi' \nb\\
&&~~~~~~~~~~ + \frac{1}{2a} \left[3 \hat{A} \hat{J_A} \psi'-\hat{A} (\delta J_A'+3 {\cal{H}} \delta J_A)\right]\nb\\
&& ~~~~~~~~~~  + \phi \hat{A} \hat{J}_\varphi- \hat{J}_\varphi \delta A\Big\}=0,\\
\lb{eq7h}
&& (v + B)' + \left(1 - 3c_{s}^{2}\right)(v +B) + \phi \nb\\
&& ~~~~~~ + \frac{1}{\hat{\rho} + \hat{p}}\left(\delta {\cal{P}}+\frac{2}{3}\partial^2 \Pi\right) = \frac{1}{2a(\hat{\rho} + \hat{p})}\Big[\hat{J}_{A}\delta{A}\nb\\
&& ~~~~~~ - 2a\hat{A}\hat{J}_{\varphi} \left(1 + c_{s}^{2}\right)\Big],
\eqn
where  $q \equiv - a(\bar{\rho}+\bar{p})(v + B)$ \cite{WM}, and $c_{s}^{2}$ denotes the adiabatic speed of sound, defined as
\bq
\lb{eq7i}
c_{s}^{2} \equiv \frac{\hat{p}'}{\hat{\rho}'}. 
\eq

It is always useful to compare the above set of field equations with those given in GR. First, because of the presence of the gauge field $A$ and the Newtonian prepotential $\varphi$,
here we have two extra equations, Eqs.(\ref{eq7a}) and (\ref{eq7b}), which are   absent in GR. As shown in Sec. IV, it is exactly  Eq.(\ref{eq7b}) in the vacuum that eliminates the
spin-0 gravitons. The momentum constraint  (\ref{eq7c}) reduces to that of GR given by Eq.(8.17) in \cite{MW09} where $\lambda = 1$.
Considering the gauge choice of Eq.(\ref{eq7.0}), the Hamiltonian constraint (\ref{eq7d}) reduces to Eq.(8.16) of \cite{MW09} for $\lambda = 1$ and $ \beta_{i} = 0$, as expected.
The same is true for the dynamical equations (\ref{eq7e}) and (\ref{eq7f}) and the conservation law of momentum (\ref{eq7h}), which will reduce, respectively,  to Eqs.(8.27),  (8.28)
and (8.33) given in \cite{MW09} for $\lambda = 1,\;
\beta_{i} =  \gamma_{2} = \gamma_{3} = \hat{A} = \delta{A} = 0$. However, because of the  foliation-preserving diffeomorphisms Diff($M, \; {\cal{F}}$) (\ref{1.4}), the conservation law
of energy (\ref{eq7g})  now takes an integral form. A direct consequence of it is that the gauge-invariant curvature perturbations
\bq
\lb{eq7j}
\zeta \equiv - \psi - {\cal{H}}\frac{\delta\rho}{\hat{\rho}'},
\eq
is not necessarily conserved on large scales even the perturbations are adiabatic \cite{WM,WWM}. In contrast, it was shown that
$\zeta$ is conserved on large scales for adiabatic perturbations in any theory of relativistic gravity, as long as the conservation law of energy holds locally
\cite{WMLL}.  Note that $\zeta$ defined here should not be confused with that introduced in the action
(\ref{action}).

\subsection{Vector Perturbations}

For the vector perturbations, we have
\bqn
\delta N&=& 0,\;\;\; \delta N^i=-S^i, \nb\\
 \delta g_{ij} &=& 2a^2 (\partial_{(i} F_{j)},\nb\\
\delta A&=&   \delta \varphi =0,
\eqn
while the corresponding matter perturbations are given by
\bqn
&&\delta J^i=\frac{1}{a^2} q^i,\;\;\delta J^t=0,\;\; \nb\\
&& \delta \tau_{ij}=2 a^2 (\Pi_{(i,j)}+\hat{p} F_{(i,j)}),
\eqn
where
\bq
\partial_i q^i=0=\partial_i \Pi^i.
\eq
Then, one finds that
\bqn
&&\delta K_{ij}=-a (F'_{(i,j)}+2{\cal{H}}F_{(i,j)}+S_{(i,j)}),\nb\\
&&\delta \Gamma_{ij}^k= \partial_{i}\partial_j F^{k},\nb\\
&&\delta R_{ij}= \delta {\cal{L}}_K=\delta {\cal{L}}_V = 0,\nb\\
&&\delta F_{ij}= - 2 a^2 \Lambda F_{(i,j)} .
\eqn
Hence, to linear order, the momentum constraint (\ref{mom}) gives
\bqn\label{ver1}
\partial^2 (F_i'+S_i)=16 \pi G a q_i,
\eqn
and the dynamical equation (\ref{dyn}) yields,
\bqn\label{ver2}
(F'_{(i,j)}+S_{(i,j)})'+2{\cal{H}} (F'_{(i,j)}+S_{(i,j)})=16 \pi G a^2 \Pi_{(i,j)}.\nb\\
\eqn
The conservation law of energy (\ref{energy conservation}) does not give new constraint,  while the conservation of
momentum (\ref{mom conservation}) yields,
\bqn
q_i'+3{\cal{H}}q_i=a\partial^2 \Pi_i.
\eqn
However, this equation is not independent, and can be obtained from Eqs. (\ref{ver1}) and (\ref{ver2}).

It is interesting to note that the vector perturbations given above are precisely the same as those presented in \cite{Wang}, in which   the projectability
condition $N = N(t)$ was assumed, but without the additional U(1) symmetry. Although one would expect some differences,
 because of the presence of the vector field $a_{i}$ defined by Eq.(\ref{vector}), a closer examination shows it is not, this is simply because  $a_{i}$ is made of $N$, and
perturbations of $\delta{N}$, as well as of $\delta{A}$ and $\delta{\varphi}$,  have no contributions  to the vector  perturbations.

\subsection{Tensor Perturbations}
The cosmological tensor perturbations are given by
\bq
\delta g_{ij}=a^2 H_{ij},\;\;\delta N^i =0,\;\; \delta N = \delta A=   \delta \varphi =0,
\eq
while the corresponding matter perturbations are given by
\bq
\delta \tau_{ij}=a^2 (\Pi_{ij}+\hat{p} H_{ij}),\;\;\;
\delta J^t=0=\delta J^i,
\eq
where
\bqn
\Pi^{i}_{\;\; i}=0,\;\;\; \partial^j \Pi_{ij} = 0.
\eqn
Then, one finds  that
\bqn
\delta K_{ij}&=&- a ({\cal{H}}H_{ij}+\frac{1}{2}H'_{ij}),\nb\\
\delta \Gamma_{ij}^k&=&\frac{1}{2} (\partial_i H^k_j+\partial_j H^k_i-\partial^k H_{ij}),\nb\\
\delta R_{ij}&=&-\frac{1}{2} \partial^2 H_{ij},\;\;\;\;\delta R=0,\nb\\
\delta \pi^{ij} &=&-\frac{1}{a^3} \Big[(1-3\lambda){\cal{H}}H^{ij}-\frac{1}{2} H'^{ij}\Big]\nb\\
\delta F_{ij}&=&-a^2 \Lambda H_{ij}+\frac{1}{2}\partial^2 H_{ij}-\frac{\gamma_3}{2a^2\zeta^2}\partial^4H_{ij}\nb\\
&&+\frac{\gamma_5}{2a^4\zeta^4}\partial^6 H_{ij}.
\eqn
In this case, all the constraints and equations are satisfied identically, except for  the dynamical one (\ref{dyn}), which  gives,
\bqn
\lb{tensor}
&&H_{ij}''+2 {\cal{H}} H_{ij}'-\left(1-\frac{\hat{A}}{a}\right)\partial^2 H_{ij}\nb\\
&&+ \frac{\gamma_3}{a^2\zeta^2}\partial^4H_{ij}- \frac{\gamma_5}{a^4\zeta^4}\partial^6 H_{ij}
=16\pi G a^2 \Pi_{ij}.
\eqn
When $\hat{A} = 0$, it reduces precisely to the one given in \cite{Wang} for the case  without the additional U(1) symmetry.

This completes the general descriptions for the scalar, vector, and tensor perturbations in our current setup.

\section{Conclusions}

There are two major variants of Horava-Lifshitz gravity, which have the potential to solve all the problems found so far. One  is the HMT generalization \cite{HMT}, which adopts the projectability condition and introduces a gauge filed $A$ and a Newtonian prepotential $\varphi$ to eliminate the spin-0 gravitons.  Another setup is due to BPS \cite{BPS}, who abandoned the projectibility condition and improved the IR limit of the theory by introducing the vector field $a_i$, defined by Eq.(\ref{vector}). However, the inclusion of $a_i$ gives rise to a proliferation of independent coupling constants.

In this paper, we have considered a new generalization of Horava-Lifshitz gravity without projectability condition but with detailed balance condition softly breaking. In order to reduce the number of independent coupling constants of the non-projectability Horava-Lifshitz gravity, in Sect II we have imposed the ``generalized'' detailed balance condition, so that the number of the  independent coupling constants is dramatically reduced.
However, for the theory to have a healthy IR limit, we have allowed the     detailed balance condition   to be broken softly,  by adding all the low dimensional relevant terms. Even with those relevant terms, the number of independently coupling constants is still significantly reduced from more than 70 to 15.

However, it was found that this is not sufficient, because the detailed balance condition, even allowed to be broken softly, still prevents the existence of the sixth-order spatial derivative terms in the gravitational sector.
As a result,  the theory is not power-counting  renormalizable and the strong coupling problem cannot be solved. To resolve this problem, in Sec III, we have extended the original foliation-preserving diffeomorphism symmetry to include a local $U(1)$ symmetry, i.e., $U(1) \ltimes {\mbox{Diff}}(M, \; {\cal{F}})$. With this enlarged symmetry, in Sec IV, we have shown explicitly that the spin-0 gravitons are eliminated, and thus all the problems related to them   in the gravitational sector disappear, including the ghost, instability, strong coupling, and different speeds.

In Sec V, we have considered the coupling of  a scalar field   to the theory, and found  that in the Minkowski background  it is stable in the   both IR and UV, and becomes strong coupling for processes with energy higher than $\Lambda_\omega \equiv (M_{pl}/c_1)^{3/2}M_{pl}|c_\psi|^{5/2}$. However,   this problem can be easily cured by introducing a new energy scale $M_*$, so that $M_*< \Lambda_\omega$, where $M_*$ denotes the suppression energy scale of the sixth order derivative terms of the theory.

In Sec VI,  we have considered  cosmological applications, and found that the FRW universe  is necessarily flat in such a setup. In Sec VII,  we have studied  the scalar, vector, and tensor perturbations, and derived the general field equations  for each kind of these perturbations. For the scalar perturbations, we have written the field equations closely following those given in GR \cite{MW09}, so one can see clearly the differences between these two theories. For the vector  perturbations, they are the same as those given in \cite{Wang} for the case with the projectability condition (\ref{1.6}) but with only  the  foliation-preserving
diffeomorphisms (\ref{1.4}), while for the tensor perturbations,  the only difference is the term proportional to $\hat{A}$ in Eq.(\ref{tensor}).
This is simply because that the lapse function ${N}$, the gauge field ${A}$ and the Newtonian prepotential ${\varphi}$  all transform like  scalars under the spatial coordinate transformations
of Eq. (\ref{1.4}), and hence their linear perturbations have no contributions to the vector and tensor perturbations of both gravitational and matter sectors. 

 It would be very interesting to apply those formulas to the studies of the early universe as well as to the ones of  its large-scale structure formation.

~\\{\bf Acknowledgements:}   We would like to thank Kai Lin for his valuable discussions and comments.
This work was supported in part by DOE  Grant, DE-FG02-10ER41692 (AW);  NSFC No. 11173021   (AW); NSFC No. 11075141 (AW); NSFC No. 11005165   (FWS);  NSFC No. 11047008 (QW, TZ); and
NSFC No. 11105120  (TZ).

\section*{AppendixA: $F_V,\;F_\varphi,\;F_\lambda,\;F_{ij},\;F_{ij}^a\;$ and $F_{ij}^\varphi$}
\renewcommand{\theequation}{A.\arabic{equation}} \setcounter{equation}{0}

  $F_V,\;F_\varphi $ and $F_\lambda$, defined in Eq.(\ref{hami}),  are given by, 
\bqn\label{a1}
F_V &=&  \beta_0 ( 2 a_i^i + a_i a^i) - \frac{\beta_1}{\zeta^2} \Bigg[3 (a_i a^i)^2 + 4 \nabla_i (a_k a^k a^i)\Bigg]\nb\\
    &&  +\frac{\beta_2}{\zeta^2}\Bigg[ (a_i^i)^2 + \frac{2}{N} \nabla^2 (N a_k^k)\Bigg]\nb\\
    && - \frac{\beta_3}{\zeta^2}\Bigg[(a_i a^i) a_j^j + 2 \nabla_i (a_j^j a^i) - \frac{1}{N} \nabla^2 (N a_i a^i)\Bigg]\nb\\
    &&+ \frac{\beta_4}{\zeta^2}\Bigg[a_{ij} a^{ij} + \frac{2}{N} \nabla_j \nabla_i (N a^{ij})\Bigg]\nb\\
      && - \frac{\beta_5}{\zeta^2}\Bigg[R (a_i a^i) + 2 \nabla_i (R a^i)\Bigg]\nb\\
      &&- \frac{\beta_6}{\zeta^2}\Bigg[a_i a_j R^{ij} + 2\nabla_i (a_j R^{ij})
      \Bigg]\nb\\
      && +  \frac{\beta_7}{\zeta^2}\Bigg[ R a^i_i + \frac{1}{N} \nabla^2 (NR)\Bigg]\nb\\
      &&+ \frac{\beta_8}{\zeta^4}\Bigg[(\Delta a^i)^2 - \frac{2}{N} \nabla^i [\Delta (N \Delta a_i)]\Bigg],
\eqn
\bqn\label{a2}
F_\varphi &=& -  {\cal{G}}^{ij}\nabla_i \varphi \nabla_j \varphi, - \frac{2}{N} \hat{{\cal{G}}}^{ijkl} \nabla_l (N K_{ij} \nabla_k \varphi),\nb\\
        &&  - \frac{4}{3}\hat{{\cal{G}}}^{ijkl} \nabla_l (\nabla_k \varphi \nabla_i \nabla_j \varphi)\nb\\
         &&- \frac{5}{3}\hat{{\cal{G}}}^{ijkl} \Big[(a_i \nabla_j \varphi) (a_k \nabla_l \varphi)+\nabla_i(a_k\nabla_j\varphi\nabla_l\varphi)\nb\\
         && +\nabla_k(a_i\nabla_j\varphi\nabla_l\varphi)\Big]\nb\\
         &&+ \frac{2}{3}\hat{{\cal{G}}}^{ijkl}\Big[a_{ik} \nabla_j \varphi \nabla_l \varphi + \frac{1}{N} \nabla_i\nabla_k (N\nabla_j\varphi\nabla_l\varphi)\Big],\nb\\
\eqn
\bqn\lb{a3a}
F_\lambda &=& (1-\lambda) \Bigg\{(\nabla^2 \varphi + a_i \nabla^i \varphi)^2 - \frac{2}{N} \nabla_i (NK\nabla^i \varphi)\nb\\
           && - \frac{2}{N} \nabla_i \Big[N (\nabla^2 \varphi + a_i \nabla^i \varphi) \nabla^i \varphi\Big]\Bigg\}\label{a3}.
\eqn

 $\left(F_n\right)_{ij}$, $\left(F^{a}_{s}\right)_{ij}$ and $\left(F^{\varphi}_{q}\right)_{ij}$, defined in Eq.(\ref{tauij}),  are given, respectively, by
\bqn\lb{a4}
(F_0)_{ij} &=& -\frac{1}{2}g_{ij},\nb\\
(F_1)_{ij} &=& R_{ij}-\frac{1}{2}Rg_{ij}+\frac{1}{N}(g_{ij}\nabla^2 N-\nabla_j\nabla_i N),\nb\\
(F_2)_{ij} &=& -\frac{1}{2}g_{ij}R^2+2RR_{ij}\nb\\
             &&  +\frac{2}{N}\left[g_{ij}\nabla^2(NR)-\nabla_j\nabla_i(NR)\right],\nb\\
(F_3)_{ij} &=& -\frac{1}{2}g_{ij}R_{mn}R^{mn}+2R_{ik}R^k_{j}\nb\\
             &&  - \frac{1}{N}\Big[2\nabla_k\nabla_{(i}(NR_{j)}^k)\nb\\
             &&  -\nabla^2(NR_{ij})- g_{ij}\nabla_m\nabla_n(NR^{mn})\Big], \nb\\
(F_4)_{ij} &=&  -\frac{1}{2}g_{ij}R^3+3R^2R_{ij}\nb\\
             &&   +\frac{3}{N}\Big(g_{ij}\nabla^2-\nabla_j\nabla_i\Big)(NR^2),\nb\\
(F_5)_{ij} &=& -\frac{1}{2}g_{ij}RR_{mn}R^{mn}\nb\\
             &&+R_{ij}R_{mn}R^{mn}+2RR_{ik}R^k_{j}\nb\\
             &&  +\frac{1}{N}\Big[g_{ij}\nabla^2(NR_{mn}R^{mn})\nb\\
             &&-\nabla_j\nabla_i(NR_{mn}R^{mn})\nb\\
             &&  +\nabla^2(NRR_{ij})+g_{ij}\nabla_m\nabla_n(NRR^{mn})\nb\\
             &&  -2\nabla_m\nabla_{(i}(R^m_{j)}NR)\Big], \nb\\
(F_6)_{ij} &=& -\frac{1}{2}g_{ij}R^m_nR^n_lR^l_m+3R_{mn}R_{mi}R_{nj}\nb\\
             &&  +\frac{3}{2N}\Big[g_{ij}\nabla_m\nabla_n(NR^m_aR^{na}) \nb\\
             &&+ \nabla^2(NR_{mi}R^m_j)
              -2\nabla_m\nabla_{(i}(NR_{j)n}R^{mn})\Big], \nb\\
(F_7)_{ij} &=& -\frac{1}{2}g_{ij}R\nabla^2R+R_{ij}\nabla^2R+R\nabla_i\nabla_j R\nb\\
             &&  +\frac{1}{N}\Big[g_{ij}\nabla^2(N\nabla^2R)-\nabla_j\nabla_i(N\nabla^2R)\nb\\
             &&+R_{ij}\nabla^2(NR)
              +g_{ij}\nabla^4(NR)-\nabla_j\nabla_i (\nabla^2 (NR))\nb\\
             &&  - \nabla_{(j}(NR\nabla_{i)}R)+\frac{1}{2}g_{ij}\nabla_k(NR\nabla^kR)\Big], \nb\\
(F_8)_{ij} &=& -\frac{1}{2}g_{ij}(\nabla_mR_{nl})^2 + 2 \nabla^mR^n_i\nabla_mR_{nj}\nb\\
             &&  +\nabla_iR^{mn}\nabla_jR_{mn}+\frac{1}{N}\Big[2 \nabla_n\nabla_{(i}\nabla_m(N\nabla^mR^n_{j)})\nb\\
             &&  -\nabla^2\nabla_m(N\nabla^mR_{ij})-g_{ij}\nabla_n\nabla_p\nabla_m(N\nabla^mR^{np})\nb\\
             &&  -2\nabla_m(NR_{l(i}\nabla^mR^l_{j)})-2\nabla_n(NR_{l(i}\nabla_{j)}R^{nl})\nb\\
             &&  +2\nabla_k(NR^k_l\nabla_{(i}R^l_{j)})\Big], \nb\\
(F_9)_{ij} &=& -\frac{1}{2} g_{ij} a_k G^k+\frac{1}{2} \Big[a^k R_{k((j} \nabla_{i)} R + a_{(i} R_{)jk} \nabla^k R\Big]\nb\\
           &&-a_kR_{mi}\nabla_jR^{mk}-a^kR_{in}\nabla^nR_{jk}-a_iR^{km}\nabla_mR_{kj}\nb\\
           &&-\frac{3}{8}a_{(i}R\nabla_{j)}R+\frac{3}{8}\Bigg\{R\nabla_k(Na^k)R_{ij}\nb\\
           &&+g_{ij}\nabla^2\Big[R\nabla_k(Na^k)\Big]-\nabla_i\nabla_j\Big[R\nabla_k(Na^k)\Big]\Bigg\}\nb\\
           &&+\frac{1}{4N} \Bigg\{- \frac{1}{2}\nabla^m \Big[\nabla_i (Na_j\nabla_m R+Na_m\nabla_j R)\nb\\
           &&+\nabla_j (Na_i\nabla_m R+Na_m\nabla_i R)\Big]\nb\\
           &&+\nabla^2 (N a_{(i}\nabla_{j)}R)+g_{ij} \nabla^m\nabla^n (Na_m\nabla_nR)\nb\\
           && +\nabla^m\Big[\nabla_i(Na_k\nabla_jR^k_m+Na_k\nabla_mR^k_j)\nb\\
           &&+\nabla_j(Na_k\nabla_iR^k_m+Na_k\nabla_mR^k_i)\Big]\nb\\
           &&-2\nabla^2(Na_k\nabla_{(i} R^k_{j)})-2g_{ij}\nabla^m\nabla^n(Na_k\nabla_{(n}R_{m)}^k)\nb\\
           &&- \nabla^m \Big[\nabla_i\nabla_p(Na_jR_m^p+Na_mR_j^p)\nb\\
           &&+\nabla_j\nabla_p(Na_iR_m^p+Na_mR_i^p)\Big]\nb\\
           &&+2\nabla^2\nabla_p(Na_{(i}R_{j)}^p)\nb\\
           && +2g_{ij}\nabla^m\nabla^n \nabla^p(Na_{(n}R_{m)p})\Bigg\}, \nb\\
\eqn
%
\bqn\lb{a5}
(F_0^a)_{ij} &=&  -\frac{1}{2} g_{ij} a^k a_k +a_i a_j, \nb\\
(F_1^a)_{ij} &=&  -\frac{1}{2} g_{ij} (a_k a^k)^2 + 2 (a_k a^k) a_i a_j,\nb\\
(F_2^a)_{ij} &=&  -\frac{1}{2} g_{ij} (a_k^{\;\; k})^2 + 2 a_k^{\;\; k} a_{ij}\nb\\
             &&   - \frac{1}{N} \Big[2 \nabla_{(i} (N a_{j)} a_k^{\;\; k}) - g_{ij} \nabla^l (a_l N a_k^{\;\; k})\Big],\nb\\
(F_3^a)_{ij} &=&   -\frac{1}{2} g_{ij} (a_k a^k) a_l^{\;\; l} + a^k_{\;\;k} a_ia_j + a_k a^k a_{ij}\nb\\
             &&   - \frac{1}{N} \Big[ \nabla_{(i} (N a_{j)} a_k a^k) - \frac{1}{2} g_{ij} \nabla^l (a_l N a_ka^k)\Big],\nb\\
(F_4^a)_{ij} &=&  - \frac{1}{2} g_{ij} a^{mn} a_{mn} + 2a^k_{\;\; i} a_{kj} \nb\\
             &&   - \frac{1}{N} \Big[\nabla^k (2 N a_{(i} a_{j)k} - N a_{ij} a_k)\Big], \nb\\
(F_5^a)_{ij} &=&  -\frac{1}{2} g_{ij} (a_k a^k ) R + a_i a_j R + a^k a_k R_{ij} \nb\\
              &&  + \frac{1}{N} \Big[ g_{ij} \nabla^2 (N a_k a^k) - \nabla_i \nabla_j (N a_k a^k)\Big], \nb\\ 
(F_6^a)_{ij} &=&   -\frac{1}{2} g_{ij} a_m a_n R^{mn} +2 a^m R_{m (i} a_{j)} \nb\\
              &&  - \frac{1}{2N} \Big[ 2 \nabla^k \nabla_{(i} (a_{j)} N a_k) - \nabla^2 (N a_i a_j) \nb\\
               && - g_{ij} \nabla^m \nabla^n (N a_m a_n)\Big], \nb\\ 
(F_7^a)_{ij} &=&  -\frac{1}{2} g_{ij} R  a_k^{\;\; k} +  a_k^{\;\; k} R_{ij} + R a_{ij} \nb\\
             &&   + \frac{1}{N} \Big[ g_{ij} \nabla^2 (N  a_k^{\;\; k}) - \nabla_i \nabla_j (N  a_k^{\;\; k}) \nb\\
             &&  - \nabla_{(i} (N R a_{j)}) + \frac{1}{2} g_{ij} \nabla^k (N R a_k)\Big], \nb\\
(F_8^a)_{ij} &=&  -\frac{1}{2} g_{ij} (\Delta a_k)^2 + (\Delta a_i) (\Delta a_j) + 2 \Delta a^k \nabla_{(i} \nabla_{j)} a_k \nb\\
             &&   + \frac{1}{N} \Big[\nabla_k [a_{(i} \nabla^k (N \Delta a_{j)}) + a_{(i} \nabla_{j)} (N \Delta a^k)\nb\\
             &&   - a^k \nabla_{(i} (N \Delta a_{j)}) + g_{ij} N a^{l k} \Delta a_l  - N a_{ij} \Delta a^k ]\nb\\
             &&  -  2 \nabla_{(i} (N a_{j)k} \Delta a^k)\Big],
\eqn
%
\bqn\lb{a6}
(F_1^\varphi)_{ij} &=&     -\frac{1}{2} g_{ij} \varphi {\cal{G}}^{mn}K_{mn}\nb\\
                   &&  + \frac{1}{2\sqrt{g} N}  \partial_t (\sqrt{g} \varphi {\cal{G}}_{ij}) -
                                                                                2 \varphi K_{(i}^l R_{j) l} \nb\\
                   &&       + \frac{1}{2} \varphi (K R_{ij} + K_{ij} R - 2 K_{ij} \Lambda_g ) \nb\\
                   &&      + \frac{1}{2N} \bigg\{2 {\cal{G}}_{k(i} \nabla^k (N_{j)} \varphi)-{\cal{G}}_{ij} \nabla^k (\varphi N_k) \nb\\
                   &&      +  g_{ij}\nabla^2 (N \varphi K) - \nabla_i \nabla_j (N \varphi K) \nb\\
                   &&+ 2  \nabla^k \nabla_{(i} (K_{j)k} \varphi N),\nb\\
                   &&      - \nabla^2 (N \varphi K_{ij}) - g_{ij} \nabla^k \nabla^l (N \varphi K_{kl})\bigg\}, \nb\\
(F_2^\varphi)_{ij} &=&   - \frac{1}{2}g_{ij} \varphi {\cal{G}}^{mn} \nabla_m\nabla_n \varphi \nb\\
                   &&     - 2 \varphi \nabla_{(i} \nabla^k R_{j)k} + \frac{1}{2} \varphi (R- 2 \Lambda_g) \nabla_i \nabla_j \varphi \nb\\
                   &&     - \frac{1}{N} \bigg\{- \frac{1}{2} (R_{ij} + g_{ij} \nabla^2 - \nabla_i \nabla_j )(N \varphi \nabla^2 \varphi)\nb\\
                   &&      - \nabla_k \nabla_{(i} (N \varphi \nabla^k \nabla_{j)} \varphi) + \frac{1}{2}\nabla^2 (N \varphi \nabla_i \nabla_j \varphi) \nb\\
                   &&      + \frac{g_{ij}}{2} \nabla^k \nabla^l ( N \varphi \nabla_k \nabla_l \varphi)\nb\\
                   &&      - {\cal{G}}_{k (i} \nabla^k (N \varphi \nabla_{j)}\varphi ) + \frac{1}{2} {\cal{G}}_{ij} \nabla^k (N \varphi \nabla_k \varphi)\bigg\},\nb\\
(F_3^\varphi)_{ij} &=&    - \frac{1}{2}g_{ij} \varphi {\cal{G}}^{mn} a_m\nabla_n \varphi \nb\\
                   &&    -\varphi ( a_{(i} R_{j)k} \nabla^k \varphi + a^k R_{k(i} \nabla_{j)} \varphi)\nb\\
                   &&      + \frac{1}{2} (R - 2 \Lambda_g)  \varphi a_{(i} \nabla_{j)} \varphi \nb\\
                   &&      - \frac{1}{N}\bigg\{ - \frac{1}{2} (R_{ij} + g_{ij} \nabla^2 - \nabla_i \nabla_j ) (N \varphi a^k \nabla_k \varphi)\nb\\
                   &&     - \frac{1}{2} \nabla^k \Big[  \nabla_{(i} (\nabla_{j)} \varphi N \varphi)+ \nabla_{(i} (a_{j)} \varphi N \nabla_k \varphi) \Big] \nb\\
                   &&     + \frac{1}{2}\nabla^2 (N \varphi a_{(i} \nabla_{j)} \varphi) \nb\\
                   &&+ \frac{g_{ij}}{2} \nabla^k \nabla^l (N \varphi a_k \nabla_l \varphi)\bigg\}, \nb\\ 
(F_4^\varphi)_{ij} &=&    - \frac{1}{2}g_{ij} \hat{{\cal{G}}}^{mnkl}K_{mn} a_{(k}\nabla_{l)}\varphi \nb\\
                   &&    + \frac{1}{2 \sqrt{g} N} \partial_t [\sqrt{g} {\cal{G}}_{ij}^{\;\;k l} a_{(l} \nabla_{k)} \varphi] \nb\\
                   &&      + \frac{1}{2N} \nabla^l \Big[a_l N_{(i} \nabla_{j)} \varphi +  N_{(i} a_{j)} \nabla_l  \varphi \nb\\
                   &&      -  N_l a_{(i} \nabla_{j)} \varphi + 2 g_{ij} N_l a^k \nabla_k \varphi \Big] \nb\\
                   &&      + \frac{1}{N} \nabla_{(i} (N N_{j)} a^k \nabla_k \varphi)\nb\\
                   &&      + a^k K_{k(i} \nabla_{j)} \varphi + a_{(i} K_{j)k} \nabla^k \varphi \nb\\
                   &&      - K a_{(i} \nabla_{j)} \varphi - K_{ij} a^k \nabla_k \varphi, \nb\\
(F_5^\varphi)_{ij} &=&     -\frac{1}{2} g_{ij} \hat{{\cal{G}}}^{mnkl}[a_{(k}\nabla_{l)}\varphi][\nabla_m\nabla_n\varphi]\nb\\
                    && -a_{(i} \nabla^k \nabla_{j)} \varphi \nabla_k \varphi - a_k \nabla^k \nabla_{(i} \varphi \nabla_{j)}\varphi\nb\\
                   &&     + a_{(i} \nabla_{j)} \varphi \nabla^2 \varphi + a^k \nabla_k \varphi \nabla_i\nabla_j \varphi \nb\\
                   &&     + \frac{1}{2N} \bigg\{ \nabla^k (N \varphi a_k \nabla_i \varphi \nabla_j \varphi) \nb\\
                   &&   - 2 \nabla_{(i} (N \nabla_{j)} \varphi a^k
                          \nabla_k \varphi) \nb\\
                   &&     +g_{ij} \nabla^l (\nabla_l \varphi a^k \nabla_k \varphi)\bigg\}, \nb\\
(F_6^\varphi)_{ij} &=&   - \frac{1}{2} g_{ij}\hat{{\cal{G}}}^{mnkl} [a_{(m}\nabla_{n)}\varphi][a_{(k}\nabla_{l)}\varphi]\nb\\
                     &&-\frac{1}{2} (a^k \nabla_i \varphi- a_i \nabla^k \varphi) (a_k \nabla_j \varphi - a_j \nabla_k \varphi), \nb\\
(F_7^\varphi)_{ij} &=&   -\frac{1}{2} g_{ij}   \hat{{\cal{G}}}^{mnkl} [\nabla_{(n}\varphi][a_{m)(k}][\nabla_{l)}\varphi]\nb\\
                    && -\frac{1}{2} a_k^{\;\; k} \nabla_i \varphi \nabla_j \varphi - \frac{1}{2} a_{ij} \nabla^k \varphi \nabla_k \varphi \nb\\
                     &&     +  a^k_{(i} \nabla_{j)} \varphi \nabla_k \varphi - \frac{1}{2N}\bigg \{- \nabla_{(i} (N a_{j)} \nabla_k \varphi \nabla^k \varphi) \nb\\ &&+ \nabla^k (N a_{(i} \nabla_{j)} \varphi \nabla_k \varphi)\nb\\
                   &&      + \frac{g_{ij}}{2} \nabla^k (N a_k \nabla^m \varphi \nabla_m \varphi) \nb\\
                   &&- \frac{1}{2}\nabla^k (N a_k \nabla_i \varphi \nabla_j \varphi)\bigg\}, \nb\\
(F_8^\varphi)_{ij} &=&    - \frac{1}{2} g_{ij} (\nabla^2 \varphi+a_k\nabla^k\varphi)^2\nb\\
                   &&    -2 (\nabla^2 \varphi + a_k \nabla^k \varphi) (\nabla_i \nabla_j \varphi + a_i \nabla_j \varphi ) \nb\\
                   &&      -\frac{1}{N} \bigg \{ - 2 \nabla_{(j} [N \nabla_{i)} \varphi (\nabla^2 \varphi + a_k \nabla^k \varphi)] \nb\\
                   &&       + g_{ij} \nabla^l [N (\nabla^2 \varphi + a_k \nabla^k \varphi) \nabla_l \varphi]\bigg\}, \nb\\
(F_9^\varphi)_{ij} &=&    - \frac{1}{2} g_{ij}(\nabla^2 \varphi+a_k\nabla^k\varphi)K \nb\\
                    && - (\nabla^2 \varphi + a_k \nabla^k \varphi) K_{ij} \nb\\
                   &&     - (\nabla_i \nabla_j \varphi + a_i \nabla_j \varphi ) K \nb\\
                  &&     +\frac{1}{2 \sqrt{g} N}  \partial_t [\sqrt{g} (\nabla^2 \varphi + a_k \nabla^k \varphi) g_{ij}] \nb\\
                    &&      - \frac{1}{N}\bigg\{ - \nabla_{(j} [ N_{i)} (\nabla^2 \varphi + a_k \nabla^k \varphi)] \nb\\
                    &&     + \frac{1}{2} g_{ij} \nabla^l [ N_l (\nabla^2 \varphi + a_k \nabla^k \varphi) ]\nb\\
                    &&    -  \nabla_{(j} (N K \nabla_{i)} \varphi) + \frac{1}{2} g_{ij} \nabla_k (N K \nabla^k \varphi)
                            \bigg\}.\nb\\
\eqn

\section*{Appendix B: Some Quantities for Scalar Perturbations}
\renewcommand{\theequation}{B.\arabic{equation}} \setcounter{equation}{0}

To first order, the $(F_s)_{ij}$ are given by
\bqn
(F_0)_{ij} &=& -\frac{1}{2} a^2 \delta_{ij} + a^2 \psi \delta_{ij},\nb\\
(F_1)_{ij} &=& -(\partial^2 \psi - \partial^2 \phi)\delta_{ij} +\partial_i \partial_j (\psi-\phi),\nb\\
(F_2)_{ij} &=& -\frac{8}{a^2} \left(\partial_i\partial_j-\delta_{ij}\partial^2\right)\partial^2 \psi,\nb\\
(F_3)_{ij} &=&  -\frac{3}{a^2} (\partial_i \partial_j - \delta_{ij}\partial^2) \partial^2 \psi,\nb\\
(F_7)_{ij} &=&   \frac{8}{a^4} (\delta_{ij} \partial^2 - \partial_i \partial_j ) \partial^4 \psi,\nb\\
(F_8)_{ij} &=&  - \frac{3}{a^4} ( \delta_{ij} \partial^6 \psi - \partial_i \partial_j \partial^4 \psi),
\eqn
and $(F_4)_{ij}    = (F_5)_{ij}  =  (F_6)_{ij} =  (F_9)_{ij} = 0$. Thus, we obtain
\bqn
\delta F_{ij}&=&  2 \Lambda a^2 \psi \delta_{ij} + \partial^2 (\psi-\phi)\delta_{ij} - \partial_i \partial_j (\psi-\phi)\nb\\
&&- \alpha_1 (\partial_i \partial_j + \delta_{ij})\partial^2 \psi.
\eqn
We also find that the only non-vanishing component of $\left(F_s^a\right)_{ij}$ is,
 \bqn
(F_7^a)_{ij}= -\frac{1}{a^2} (\partial_i \partial_j - \delta_{ij} \partial^2)\partial^2 \phi.
\eqn
In addition, we  have the following,
\bqn
\delta {\cal{G}}^{ij} = \frac{\partial^i\partial^j \psi-\delta^{ij}\partial^2 \psi}{a^4},\\
\delta ({\cal{G}}^{ij} K_{ij}) = \frac{2{\cal{H}}}{a^3} \partial^2 \psi,\\
\delta \left(\frac{1}{N} {\cal{G}}^{ijkl} \nabla_{(k} [N a_{l)} K_{ij}]\right) = \frac{2{\cal{H}}}{a^3} \partial^2 \phi.
\eqn


\end{document}